\documentclass[
aip,
 amsmath,amssymb,
 reprint,%
]{revtex4-1}

\usepackage{graphicx}% Include figure files
\usepackage{dcolumn}% Align table columns on decimal point
\usepackage{bm}% bold math
\usepackage{subcaption} % for subfigures
\usepackage{color}
\usepackage[utf8]{inputenc}
\usepackage[T1]{fontenc}
\usepackage{mathptmx}
\usepackage{etoolbox}
\usepackage{float} % for the [H] placement option

\usepackage{parskip}

\usepackage{hyperref} % for hyperlinks
\usepackage{orcidlink}

\usepackage{titlesec}
\titlespacing{\subsection}{1pt}{\parskip}{0pt}
\titlespacing{\section}{0pt}{\parskip}{0pt}

\makeatletter
\def\@email#1#2{%
 \endgroup
 \patchcmd{\titleblock@produce}
  {\frontmatter@RRAPformat}
  {\frontmatter@RRAPformat{\produce@RRAP{*#1\href{mailto:#2}{#2}}}\frontmatter@RRAPformat}
  {}{}
}%
\makeatother
\newcommand{\IV}{$I$-$V$\!}
\begin{document}

\preprint{AIP/123-QED}

\title[High-temperature domain wall current in Mg-doped lithium niobate single crystals up to 400$^\circ$C]{High-temperature domain wall current in Mg-doped lithium niobate single crystals up to 400$^\circ$C}
% Force line breaks with \\
\author{Uliana Yakhnevych}
  \email{uliana.yakhnevych@tu-clausthal.de}
\author{Marlo Kunzner}%
\affiliation{ Institut f\"ur Energieforschung und Physikalische Technologien, Technische Universit\"at Clausthal, Am Stollen 19B, 38640 Goslar, Germany
%\\This line break forced with \textbackslash\textbackslash
}%

\author{Leonard M. Verhoff\,\orcidlink{0009-0004-3358-1312}}
\affiliation{%
	Institut f\"ur Theoretische Physik and	Center for Materials Research (ZfM/LaMa),	Justus-Liebig-Universit\"at Gießen,	Heinrich-Buff-Ring 16, 35392 Gießen, Germany%\\This line break forced% with \\
}%

\author{Julius Ratzenberger\,\orcidlink{0000-0001-6896-2554}}
\affiliation{%
Institut f\"ur Angewandte Physik, Technische Universit\"at Dresden, N\"othnitzer Stra{\ss}e 61, 01187 Dresden, Germany%\\This line break forced% with \\
}%

\author{Elke Beyreuther\,\orcidlink{0000-0003-1899-603X}}
\affiliation{%
Institut f\"ur Angewandte Physik, Technische Universit\"at Dresden, N\"othnitzer Stra{\ss}e 61, 01187 Dresden, Germany%\\This line break forced% with \\
}%

\author{Michael R\"using\,\orcidlink{0000-0003-4682-4577}}
\affiliation{%
Integrated Quantum Optics,  Institute for Photonic Quantum Systems (PhOQS), Paderborn University, 33098 Paderborn, Germany%\\This line break forced% with \\
}%

\author{Simone Sanna\, \orcidlink{0000-0003-4416-0252}}
\affiliation{%
	Institut f\"ur Theoretische Physik and	Center for Materials Research (ZfM/LaMa),	Justus-Liebig-Universit\"at Gießen,	Heinrich-Buff-Ring 16, 35392 Gießen, Germany%\\This line break forced% with \\
}%

\author{Lukas M. Eng\,\orcidlink{0000-0002-2484-4158}}
\affiliation{%
Institut f\"ur Angewandte Physik, Technische Universit\"at Dresden, N\"othnitzer Stra{\ss}e 61, 01187 Dresden, Germany%\\This line break forced% with \\
}%
\affiliation{%
Dresden-W\"urzburg Cluster of Excellence - EXC 2147, Technische Universit\"at Dresden, 01062 Dresden, Germany
}%

\author{Holger Fritze\,\orcidlink{0000-0002-9910-922X}}
\affiliation{ Institut f\"ur Energieforschung und Physikalische Technologien, Technische Universit\"at Clausthal, Am Stollen 19B, 38640 Goslar, Germany
%\\This line break forced with \textbackslash\textbackslash
}%
\affiliation{Forschungszentrum Energiespeichertechnologien, Technische Universit\"at Clausthal, Am Stollen 19A, 38640 Goslar, Germany}

\date{\today}% It is always \today, today,

\begin{abstract}
Conductive ferroelectric domain walls (DWs) represent a promising topical system for the development of nanoelectronic components and devices. DWs show very different properties as compared to their bulk counterparts. Of central interest here is the domain wall current (DWC) of charged DWs in 5mol\% Mg-doped lithium niobate single crystals; in contrast to former works, we extend the DWC study here to temperatures as high as 400$^\circ$C. Both the temporal stability and the thermal activation energies of 90 - 160 meV are readily deduced from current-voltage sweeps as recorded over multiple heating cycles. Our experimental work is backed up by atomistic modelling of the DWC. The latter suggests that a large band bending renders head-to-head and tail-to-tail DWs semimetallic. These detailed investigations underline the potential to extend DWC-based nanoelectronic applications even into the so-far unexplored high-temperature regime.
\end{abstract}

\maketitle

%\

Single-crystalline lithium niobate (LN) is the $drosophila$ ferroelectric for applications based on piezoelectricity, electro-optical effects, or polarization switching. Devices include precision actuators, non-volatile ferroelectric random-access memories, electro-optical modulators, photovoltaic cells, resistive switches, and sensors for measuring vibration and magnetic fields, even at higher temperatures \cite{Poberaj2012,Waser2004}. \\
One major key factor that contributes to the versatility of LN-based devices is the ability to elegantly engineer ferroelectric domains and domain walls (DWs) into these crystals \cite{Meier2021,McCluskey2022,Kirbus2019,Geng2021,Waser2004}. Domain engineering in LN  involves the controlled creation and manipulation of DWs \cite{Shur2015}. Recently, considerable advancements have been reported for fabricating and tuning the DW properties in LN crystals, specifically also to reproducibly increase the DW conductivity (DWC) by several orders of magnitude \cite{Werner2017,Godau2017,Zhang2022,Kirbus2019} which is of clue interest to the work here. Controlling the DWC has a broad impact for various devices, such as resistive switches, diode structures, or neuromorphic computing  \cite{Zhang2021,Sharma2023,Chai2021, Suna2022}. For sensing devices, the ability to manipulate the DWC in LN offers promising prospects, provided they meet the challenging environmental conditions of i.e. high temperatures, as requested for instance in aerospace and automotive applications \cite{Turner1994}.  In fact, despite some minor hints towards a thermally-activated DWC process up to 70$^\circ$C, studies on this topic have only rarely been reported \cite{Werner2017,Zahn2024} and the high-temperature behavior and stability of domains and DWs remain unclear to date. While domains in LN endure annealing up to the Curie temperature ($T_c$) at about 1150$^\circ$C, the stability and enhanced conductivity of DWs is uncertain due to their non-equilibrium, inclined structure \cite{Godau2017}. 

By today, the electrical transport mechanism along charged ferroelectric DWs in LN (and also in many other ferroelectrics) is understood partly, only  \cite{Beccard2023,Zahn2024,McCluskey2022,nataf20}.  While the formation of a 2-dimensional electron gas (2DEG) at strongly charged walls is proposed and experimentally investigated by many works \cite{bed18,nataf20,bec22}, DWC measurements also support a thermally-activated process at less-inclined and less-charged DWs, e.g., in LN \cite{Zahn2024}. This contradicts the metallic-nature of a 2DEG, where no signatures of thermal activation should appear. Hence, two different DWC models result: 
\begin{itemize}
    \item The band-bending concept below the Fermi-level due to strong electric field discontinuities at inclined and charged DWs, leading to the formation of a 2DEG \cite{bed18,nataf20}; 
    \item The defect-mediated hopping process of (mobile) screening charges along charged DWs, e.g. bound to polaron levels \cite{Beccard2023,xia18}. 
\end{itemize}

To clarify these issues, we (i) investigate  the DWC in LN at elevated temperatures up to 400$^\circ$C. This allows us to prove the existence of thermally-activated processes beyond room temperature, and hence to discriminate between the two mechanisms proposed above. Furthermore, (ii) we present  density functional theory (DFT) calculation of band structures of fully charged DWs in LN, in order to support whether or not the 2DEG theory accounts also for the DWC as observed here in DWs of Mg-doped LN.
 
%\section{\label{sec:methods}Method and experimental details}

%\subsubsection{\label{sec:engineerig}Engineering DWs}

In our experiments, z-cut 5mol\% Mg-doped LN wafers from Yamaju Ceramics Co., Ltd., are used for domain engineering. Every sample piece measures 200~\( \mu \)m in thickness, and has a lateral size of 5~mm $\times$ 6~mm. Single, hexagonally-shaped domains of a 110 - 300~\( \mu \)m diameter are fabricated by means of our UV-assisted electric field poling technique \cite{Haussmann2009,Godau2017}. In total, we investigated three different samples in this study, tagged as LN1, LN2, and LN3. Note that samples LN1 and LN3 carry one such poled domain denoted by LN1-a and LN3-a, respectively, while two broadly separated domains were poled into sample LN2, as denoted LN2-a and LN2-b. This variation aims at exploring potential relationships between domain size, current flow, and thermal stability. Domain formation is monitored  $in-situ$ via polarized light microscopy. Individual domains are then electrically contacted by depositing a 10-nm-thick Cr-electrode via magnetron sputtering onto both z-facets, and wired up to our electrical measuring units through conductive silver-paste contacts.  As-grown DWs typically have a very low conductivity at room temperature (RT) that then is enhanced by several orders of magnitude following the protocol of Godau et al. \cite{Godau2017,Kirbus2019}. 

DWC is investigated in the temperature range of 25-400$^\circ$C using a high-temperature setup. Here, the samples are placed on a ceramic plate with a planar platinum contact. Both are installed atop a heating module. Individual domains are contacted by a platinum tip that is precisely positioned using a 3D micromanipulator. A thermocouple is mounted at a distance of approximately 1-2~mm from the sample to control the temperature with a heating rate of 1~K/min. Current-voltage (\IV) sweeps between ±10~V are recorded using a Keithley 6517B electrometer connected to the platinum tip and the planar contact while applying voltage increments of 1.5~V. Details of the setup can be found in the supplementary material (SM, sec. S2).

%\section{\label{sec:results}Results}

%\subsubsection{\label{sec:res_I-U}I-V characteristics}

	\begin{figure}%[H]
     \includegraphics[scale=1.2]{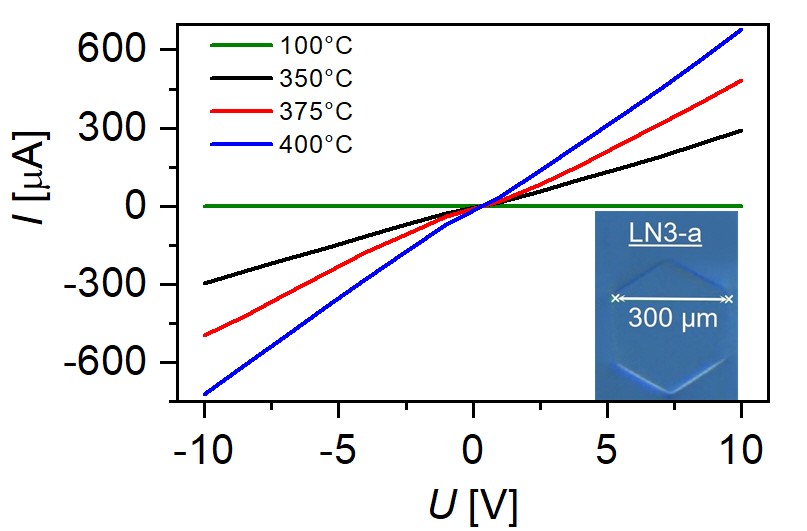}
      \caption{\IV-characteristic across the $\pm$10~V voltage bias regime of the domain wall current in domain LN3-a. The inset in depicts the DW contour of domain LN3-a, as recorded by polarized light microscopy.} 
        \label{fig:1}
\end{figure}

Fig.~\ref{fig:1} displays the typical DWC \IV-characteristics for domain LN3-a at different temperatures. Directly after domain poling illustrated also by the polarized microscopy image (see inset in Fig.~\ref{fig:1}) the DWC at RT is negligibly small across the full ±10 V range, approximately 10\( ^{-11} \)A at +7~V. A drastic increase of up to six orders of magnitude is evident after enhancement. Note that all DWs studied in this work exhibit an ohmic behavior after enhancement, while in previous works often rectifying characteristics have been reported as well \cite{Godau2017,Kirbus2019,Zahn2024}. We find absolute room temperature DWCs ranging from \(\sim\)10\( ^{-7} \) to \(\sim\)10\( ^{-6} \)A at a $+$7~V bias voltage for all DWC measurements in all domains, with differences being indicative for their effective and averaged DW inclination angle with respect to the sample c-axes \cite{Godau2017}. Notably, the DWCs studied here show no dependence on the domaim size.
Remarkably, the DWC in domain LN3-a at 400 $^\circ$C is increased at least 800 times as compared to the RT DWC at a +7~V bias voltage. The \IV-characteristics across the ±10 V voltage bias range for the DWC in all examined domains at RT, along with polarized microscopy images illustrating the domains directly after domain poling, are available in the SM (sec. S1). Additionally, the SM includes the \IV-curve across various temperatures for all studied domains (sec. S3).

%\subsubsection{\label{sec:res_I-T}Temperature dependent current}

Fig.~\ref{fig:2} displays the DWC recorded at a $+$7~V bias voltage for domains LN1-a and LN3-a in the Arrhenius-type plot, together with the DWC data from a bulk LN reference sample containing no DW (plotted in \textcolor{orange}{orange}). The DWC increases as a function of rising temperatures for all domains, including the bulk reference  (see Fig.~\ref{fig:2}, Fig. S7 and Fig. S8 in SM sec. S4). This trend is observed consistently either for increasing or decreasing temperature sweeps. Of particular interest is the comparison between investigated DWCs and bulk LN currents, as depicted in Fig.~\ref{fig:2}. The bulk LN current follows from conductivity data obtained through impedance spectroscopy \cite{Yakhnevych2024,Kofahl2024}; here, the current was calculated based on the sample thickness and average electrode size. Note the huge increase in the current magnitude between the bulk reference and domains LN1-a and LN3-a, respectively, displaying a six-order-of-magnitude difference at 250~$^\circ$C. Therefore, we can exclude any noticable bulk related contribution to the measured DWCs. 
Moreover, we notice in Fig.~\ref{fig:2} different temperature regimes (marked with solid lines and annotated in Roman numerals: I, II, III) that show a linear increase in DWC, however, with different slopes, indicating thermally-activated transport processes with different activation energies. We see from Fig.~\ref{fig:2} that different DWCs (here LN1-a and LN3-a) scatter by one order of magnitude in the absolutely measured DWC (see the DWC data for domain LN1-a and LN3-a at RT in Fig.~\ref{fig:2}). This trend persists up to approximately 200$^\circ$C, hence branding the region~I in Fig.~\ref{fig:2}. Above that threshold of 200$^\circ$C, DWC data scatter less, (region~II) and the differences gradually diminish, converging practically to the same DWC value for all analyzed DWs (in region~III). These findings suggest that the DW morphology plays no major role anymore above the 200$^\circ$C threshold.  
 
	\begin{figure}[H]%[htbp]
		    \includegraphics[scale=1.2]{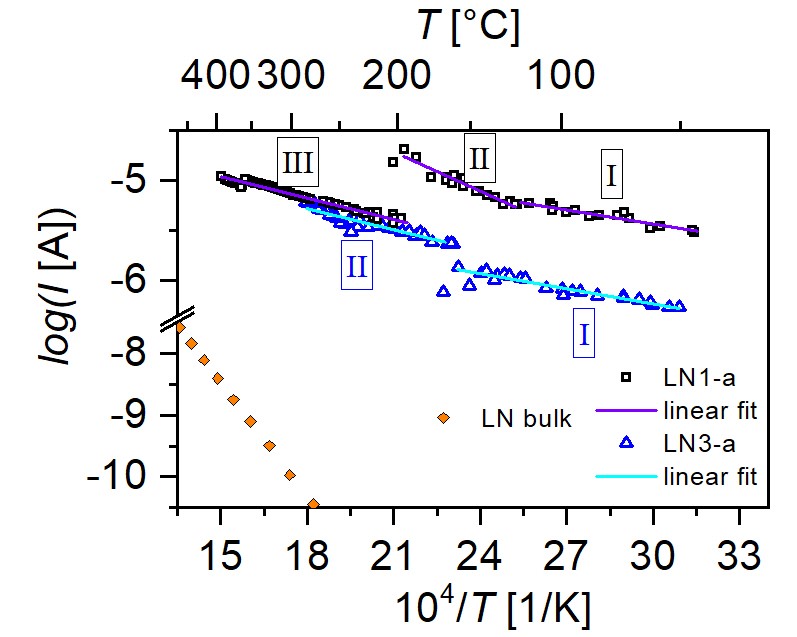}% Adjust scale if needed
     	\caption{Temperature-dependent DWC plotted for domain LN1-a (\textcolor{purple}{purple}) and LN3-a (\textcolor{blue}{blue}) at a +7~V bias voltage. For reference, the bulk LN current (in \textcolor{orange}{orange}) is displayed in the same Arrhenius plot.}
	    \label{fig:2}
	\end{figure}

%\subsubsection{\label{sec:res_EA}Activation energy}

Fig.~\ref{fig:3} presents the activation energy ($E_A$) for all investigated DWCs as extracted by fitting $I \sim \exp(E_A/k_BT)$ to the linear segments I, II, and III in Fig.~\ref{fig:2}. Here, $k_B$ represents the Boltzmann constant. The temperature intervals used for data fitting are the same ones as the ones delineated in Fig.~\ref{fig:2}. In addition, Fig.~\ref{fig:3} contains also the as-deduced $E_A$ for DW LN2-a,  depicted for both its initial DWC state (\textcolor{green}{green} solid line) and the subsequent heating cycle (\textcolor{green}{green} dashed line). Remarkably, our findings suggest that the studied domains maintain stable during repeated heating cycles. This implies a promising potential for utilizing DWC-based devices at elevated temperatures.
	
	\begin{figure}[htbp]
		\includegraphics[scale=1.2]{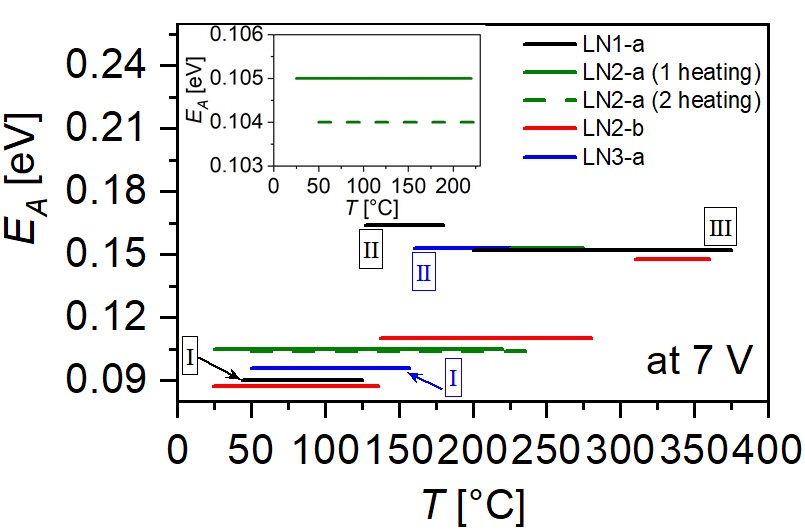}
		\caption{Activation energies $E_A$ of the DWC for the distinct temperature ranges as marked in Fig.~\ref{fig:2}. Note the reproducible $E_A$ values of DW LN2-a even after several heating cycles.}
		\label{fig:3}
	\end{figure}
 
Our observations in Fig.~\ref{fig:3} reveal two distinct activation energy levels, measuring approximately $E_A$ = 90~meV and 160~meV. The data analysis delineates the temperature ranging into two overlapping regimes: room to medium temperatures (25 - 280$^\circ$C) and temperatures above about 150$^\circ$C. Werner et al. \cite{Werner2017} and Zahn et al. \cite{Zahn2024} reported  $E_A$ values of 100~meV and 200~meV for the temperature range from -193$^\circ$C to 70$^\circ$C. They also suggested that the dominant charge carrier type for DWC in these DWs is of electronic nature, backed up by both recent Hall-transport measurements \cite{Beccard2023} and strain vs.\ conductivity investigations \cite{sin22}, specifically involving thermally activated electron-polaron hopping. We assume that the activation energy values obtained here, ranging from 90~meV to 160~meV, also indicate involvement of electron-polaron hopping processes; in fact, such $E_A$ values align well with the reported values for electron-polaron jumps in bulk lithium niobate \cite{Reichenbach2018,Koppitz1987, Schirmer2009}.

It is evident from our experiments that DWs exhibit distinct activation energies across different temperature ranges. This implies the involvement of more than one type of defect/polaron that contributes to the DWC. Significantly, the presence of varied activation energies at both room and high temperatures consistently observed within the same sample, suggests that these differences are not sample-specific but likely associated with the DW morphology, which will require further studies.

	\begin{figure}[htbp]
	\includegraphics[scale=0.23]{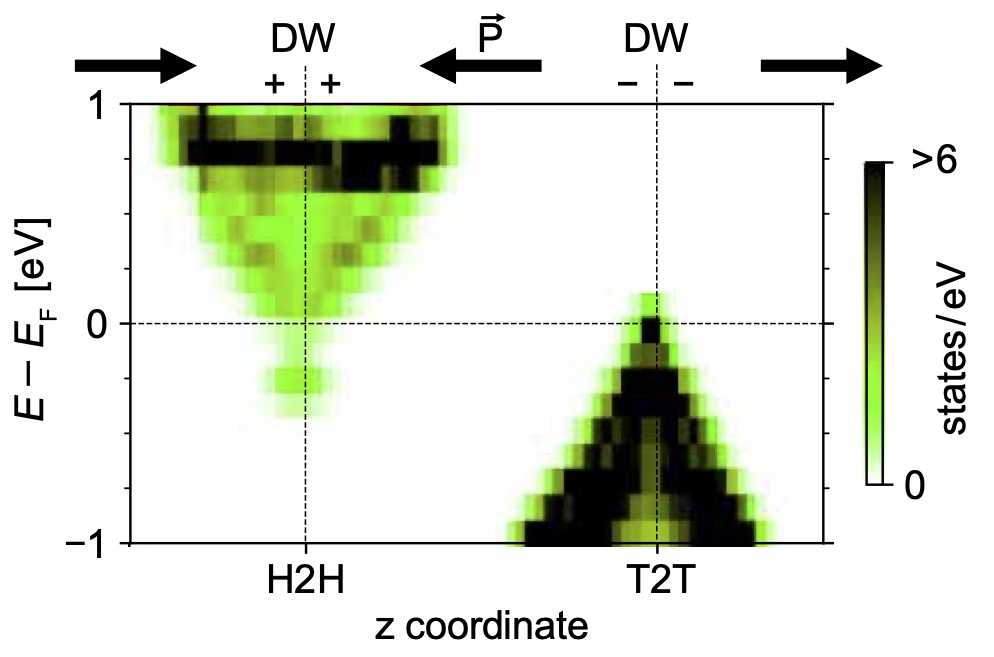}
	\caption{Local density of states (DOS) of a super-cell when modelling a fully-charged H2H and a T2T DW in LN. The direction of polarization, the position of the Fermi energy, and the sign of the polarization charges are indicated. Note the very different DOS levels for the H2H and T2T DW.}
	\label{fig:4}
\end{figure}

 In order to investigate the mechanisms behind the DWC, DWs perpendicular to the crystallographic directions X, Y and Z of stoichiometric LN are modeled within DFT in the independent particle approximation (IPA). We thereby employ the VASP software package \cite{Kresse1996,Kresse1996_2}, projector augmented-wave  (PAW) potentials implementing the  Perdew-Burke-Ernzerhof (PBE) formulation of the xc-functional \cite{Perdew1996,Perdew2008,Bloechl94} and a cutoff energy for the plane wave basis of 500~eV. Further computational details can be found in the SM (sec. S5). Structural optimization reveals that the atomic positions and the local polarization reach the value of bulk LN at a distance of about 1\,nm from the DW center in the case of X and Y DWs, and at a distance of 2\,nm in the case of the fully-charged Z walls.  This is in agreement with typical DW widths as deduced from transmission electron microscopy investigations  \cite{Gonnissen2016}. The presence of DWs deeply modifies the electronic structure, as shown by the calculated density of states (DOS) shown in the SM (Figure S9). However, while X and Y DWs reduce the fundamental bandgap from 3.4 eV to about 3.0 eV, charged Z DWs lead to an overlap of valence and conduction band, as demonstrated by the local density of states of the super-cell modelling the DWs shown in  Fig.~\ref{fig:4}. The direction of polarization and the sign of the polarization charges for the head-to-head (H2H) and the tail-to-tail (T2T) DWs are indicated as well. Interestingly,
the spontaneous polarization causes a strong band bending for both types of fully-charged DWs, so that a non-vanishing portion of electronic states belonging to the conduction (valence) band, crosses the Fermi energy in proximity to the H2H (T2T) DW \cite{Verhoff2024}. The domain walls thus become semi-metallic, similar to suggestions in other materials \cite{nataf20,bed18}. Moreover, the DOS at the H2H and T2T DWs is very different; H2H DWs feature very low DOS values at the Fermi-energy and higher DOS values about 150 meV below it. Unfortunately, due to the limited precision of the computational approach, it cannot be conclusively stated whether this gap in the H2H DWs DOS is related to the measured activation energy or not.

According to our calculations, a 2DEG is formed for fully-charged H2H and T2T DWs, that will contribute to the DWC. Nevertheless, a word of caution has to be expressed, addressing the interpretation of our DFT calculations in relation to the experiment: 

\begin{itemize}
    \item Firstly, improved methods beyond the IPA approach as used here for calculating the electronic structure, are required to quantitatively estimate the overlap between conduction and valence bands.
    \item Second, inclined DWs, such as the DW measured in the present experiments, can be considered as a sequence of non charged X and Y DWs and charged H2H and T2T walls. The first feature a less pronounced band bending and hence might not lead to the formation of a 2DEG. Whether the contribution of H2H and T2T walls within an oblique DW suffices to render the whole DW metallic is questionable. This situation will not drastically change at the temperatures discussed in this work, when vibrational contributions are expected to close the electronic bandgap by further 0.5 eV with respect to the 0\,K calculations presented here \cite{Riefer16}. 
    \item Third, the role of doping or defects, such as lithium vacancies or Mg-doping as used in the current experiment, is still unclear. They might easily lead to the formation of additional energy levels within the DWs (e.g., donor or acceptor levels) or shifts in the band structure. 
\end{itemize}

Whether or not and to which extent different mechanisms such as the 2DEG formation, any polaron accumulation and charge hopping, etc.~contribute to the DWC remains to be settled with further investigations. Nonetheless, this work demonstrates the feasibility to investigate DW band structures via first principle calculations and opens up the option of high DW currents. 
	
	%\section{\label{sec:summary}Conclusions}
	
In conclusion, our results consistently demonstrate that as temperature rises, there is a corresponding increase in the domain wall current. DWs, along with their conductivity, remain stable even upon multiple heating cycles up to about 400$^\circ$C, marking the highest temperatures documented in the existing literature. This indicates the potential for developing DW-based electronics for high-temperature applications. At temperatures up to about 400$^\circ$C, the thermally-activated nature of the DWC is confirmed, indicating that a hopping process (or similar mechanism) is also present for this temperature range. Here, no signs of saturation or an extrinsic conductivity region are found for slightly-inclined DWs. Furthermore, DWs display varied activation energies in two overlapping temperature regimes, hinting towards the involvement of more than one type of defect/polaron to conduction, consistently observed within several DWs.
	
\begin{acknowledgments}
The authors gratefully acknowledge financial support by the Deutsche Forschungsgemeinschaft (DFG) through the Research unit FOR5044 (ID: 426703838; \url{https://www.for5044.de}). We thank Thomas Gemming and Dina Bieberstein for assistance in wafer dicing, as well as Henrik Beccard for assistance in sample preparation. EB and LME thank the Würzburg-Dresden Cluster of Excellence on “Complexity and Topology in Quantum Matter” - ct.qmat (EXC 2147; ID 39085490).  Calculations for this research were conducted on the Lichtenberg high-performance computer of the TU Darmstadt and at the H\"ochstleistungrechenzentrum Stuttgart (HLRS). The authors furthermore acknowledge the computational resources provided by the HPC Core Facility and the HRZ of the Justus-Liebig-Universit\"at Gie{\ss}en.
\end{acknowledgments}

\section*{Data Availability Statement}
The data that support the findings of this study are available from the corresponding author upon reasonable request.

%\section*{References}

\nocite{*}
\bibliography{References}% Produces the bibliography via BibTeX.

\providecommand{\noopsort}[1]{}\providecommand{\singleletter}[1]{#1}%
\begin{thebibliography}{39}%
\makeatletter
\providecommand \@ifxundefined [1]{%
 \@ifx{#1\undefined}
}%
\providecommand \@ifnum [1]{%
 \ifnum #1\expandafter \@firstoftwo
 \else \expandafter \@secondoftwo
 \fi
}%
\providecommand \@ifx [1]{%
 \ifx #1\expandafter \@firstoftwo
 \else \expandafter \@secondoftwo
 \fi
}%
\providecommand \natexlab [1]{#1}%
\providecommand \enquote  [1]{``#1''}%
\providecommand \bibnamefont  [1]{#1}%
\providecommand \bibfnamefont [1]{#1}%
\providecommand \citenamefont [1]{#1}%
\providecommand \href@noop [0]{\@secondoftwo}%
\providecommand \href [0]{\begingroup \@sanitize@url \@href}%
\providecommand \@href[1]{\@@startlink{#1}\@@href}%
\providecommand \@@href[1]{\endgroup#1\@@endlink}%
\providecommand \@sanitize@url [0]{\catcode `\\12\catcode `\$12\catcode
  `\&12\catcode `\#12\catcode `\^12\catcode `\_12\catcode `\%12\relax}%
\providecommand \@@startlink[1]{}%
\providecommand \@@endlink[0]{}%
\providecommand \url  [0]{\begingroup\@sanitize@url \@url }%
\providecommand \@url [1]{\endgroup\@href {#1}{\urlprefix }}%
\providecommand \urlprefix  [0]{URL }%
\providecommand \Eprint [0]{\href }%
\providecommand \doibase [0]{http://dx.doi.org/}%
\providecommand \selectlanguage [0]{\@gobble}%
\providecommand \bibinfo  [0]{\@secondoftwo}%
\providecommand \bibfield  [0]{\@secondoftwo}%
\providecommand \translation [1]{[#1]}%
\providecommand \BibitemOpen [0]{}%
\providecommand \bibitemStop [0]{}%
\providecommand \bibitemNoStop [0]{.\EOS\space}%
\providecommand \EOS [0]{\spacefactor3000\relax}%
\providecommand \BibitemShut  [1]{\csname bibitem#1\endcsname}%
\let\auto@bib@innerbib\@empty
%</preamble>
\bibitem [{\citenamefont {Poberaj}\ \emph {et~al.}(2012)\citenamefont
  {Poberaj}, \citenamefont {Hu}, \citenamefont {Sohler},\ and\ \citenamefont
  {Günter}}]{Poberaj2012}%
  \BibitemOpen
  \bibfield  {author} {\bibinfo {author} {\bibfnamefont {G.}~\bibnamefont
  {Poberaj}}, \bibinfo {author} {\bibfnamefont {H.}~\bibnamefont {Hu}},
  \bibinfo {author} {\bibfnamefont {W.}~\bibnamefont {Sohler}}, \ and\ \bibinfo
  {author} {\bibfnamefont {P.}~\bibnamefont {Günter}},\ }\href {\doibase
  https://doi.org/10.1002/lpor.201100035} {\bibfield  {journal} {\bibinfo
  {journal} {Laser \( \& \) Photonics Reviews}\ }\textbf {\bibinfo {volume}
  {6}},\ \bibinfo {pages} {488} (\bibinfo {year} {2012})}\BibitemShut {NoStop}%
\bibitem [{\citenamefont {Waser}\ and\ \citenamefont
  {Rüdiger}(2004)}]{Waser2004}%
  \BibitemOpen
  \bibfield  {author} {\bibinfo {author} {\bibfnamefont {R.}~\bibnamefont
  {Waser}}\ and\ \bibinfo {author} {\bibfnamefont {A.}~\bibnamefont
  {Rüdiger}},\ }\href {\doibase https://doi.org/10.1038/nmat1067} {\bibfield
  {journal} {\bibinfo  {journal} {Nature Materials}\ }\textbf {\bibinfo
  {volume} {3}},\ \bibinfo {pages} {81} (\bibinfo {year} {2004})}\BibitemShut
  {NoStop}%
\bibitem [{\citenamefont {Meier}\ and\ \citenamefont
  {Selbach}(2021)}]{Meier2021}%
  \BibitemOpen
  \bibfield  {author} {\bibinfo {author} {\bibfnamefont {D.}~\bibnamefont
  {Meier}}\ and\ \bibinfo {author} {\bibfnamefont {S.}~\bibnamefont
  {Selbach}},\ }\href {\doibase https://doi.org/10.1038/s41578-021-00375-z}
  {\bibfield  {journal} {\bibinfo  {journal} {Nature Reviews Materials}\
  }\textbf {\bibinfo {volume} {7}},\ \bibinfo {pages} {157–173} (\bibinfo
  {year} {2021})}\BibitemShut {NoStop}%
\bibitem [{\citenamefont {McCluskey}\ \emph {et~al.}(2022)\citenamefont
  {McCluskey}, \citenamefont {Colbear}, \citenamefont {McConville},
  \citenamefont {McCartan}, \citenamefont {Maguire}, \citenamefont {Conroy},
  \citenamefont {Moore}, \citenamefont {Harvey}, \citenamefont {Trier},
  \citenamefont {Bangert}, \citenamefont {Gruverman}, \citenamefont {Bibes},
  \citenamefont {Kumar}, \citenamefont {McQuaid},\ and\ \citenamefont
  {Gregg}}]{McCluskey2022}%
  \BibitemOpen
  \bibfield  {author} {\bibinfo {author} {\bibfnamefont {C.~J.}\ \bibnamefont
  {McCluskey}}, \bibinfo {author} {\bibfnamefont {M.~G.}\ \bibnamefont
  {Colbear}}, \bibinfo {author} {\bibfnamefont {J.~P.~V.}\ \bibnamefont
  {McConville}}, \bibinfo {author} {\bibfnamefont {S.~J.}\ \bibnamefont
  {McCartan}}, \bibinfo {author} {\bibfnamefont {J.~R.}\ \bibnamefont
  {Maguire}}, \bibinfo {author} {\bibfnamefont {M.}~\bibnamefont {Conroy}},
  \bibinfo {author} {\bibfnamefont {K.}~\bibnamefont {Moore}}, \bibinfo
  {author} {\bibfnamefont {A.}~\bibnamefont {Harvey}}, \bibinfo {author}
  {\bibfnamefont {F.}~\bibnamefont {Trier}}, \bibinfo {author} {\bibfnamefont
  {U.}~\bibnamefont {Bangert}}, \bibinfo {author} {\bibfnamefont
  {A.}~\bibnamefont {Gruverman}}, \bibinfo {author} {\bibfnamefont
  {M.}~\bibnamefont {Bibes}}, \bibinfo {author} {\bibfnamefont
  {A.}~\bibnamefont {Kumar}}, \bibinfo {author} {\bibfnamefont {R.~G.~P.}\
  \bibnamefont {McQuaid}}, \ and\ \bibinfo {author} {\bibfnamefont {J.~M.}\
  \bibnamefont {Gregg}},\ }\href {\doibase
  https://doi.org/10.1002/adma.202204298} {\bibfield  {journal} {\bibinfo
  {journal} {Advanced Materials}\ }\textbf {\bibinfo {volume} {34}},\ \bibinfo
  {pages} {2204298} (\bibinfo {year} {2022})}\BibitemShut {NoStop}%
\bibitem [{\citenamefont {Kirbus}\ \emph {et~al.}(2019)\citenamefont {Kirbus},
  \citenamefont {Godau}, \citenamefont {Wehmeier}, \citenamefont {Beccard},
  \citenamefont {Beyreuther}, \citenamefont {Haußmann},\ and\ \citenamefont
  {Eng}}]{Kirbus2019}%
  \BibitemOpen
  \bibfield  {author} {\bibinfo {author} {\bibfnamefont {B.}~\bibnamefont
  {Kirbus}}, \bibinfo {author} {\bibfnamefont {C.}~\bibnamefont {Godau}},
  \bibinfo {author} {\bibfnamefont {L.}~\bibnamefont {Wehmeier}}, \bibinfo
  {author} {\bibfnamefont {H.}~\bibnamefont {Beccard}}, \bibinfo {author}
  {\bibfnamefont {E.}~\bibnamefont {Beyreuther}}, \bibinfo {author}
  {\bibfnamefont {A.}~\bibnamefont {Haußmann}}, \ and\ \bibinfo {author}
  {\bibfnamefont {L.}~\bibnamefont {Eng}},\ }\href {\doibase
  https://doi.org/10.1021/acsanm.9b01240} {\bibfield  {journal} {\bibinfo
  {journal} {ACS Applied Nano Materials}\ }\textbf {\bibinfo {volume} {2}},\
  \bibinfo {pages} {5787} (\bibinfo {year} {2019})}\BibitemShut {NoStop}%
\bibitem [{\citenamefont {Geng}\ \emph {et~al.}(2021)\citenamefont {Geng},
  \citenamefont {He}, \citenamefont {Qiao}, \citenamefont {Niu}, \citenamefont
  {Zhao}, \citenamefont {Xue}, \citenamefont {Bi}, \citenamefont {Mei},
  \citenamefont {Wang},\ and\ \citenamefont {Chou}}]{Geng2021}%
  \BibitemOpen
  \bibfield  {author} {\bibinfo {author} {\bibfnamefont {W.}~\bibnamefont
  {Geng}}, \bibinfo {author} {\bibfnamefont {J.}~\bibnamefont {He}}, \bibinfo
  {author} {\bibfnamefont {X.}~\bibnamefont {Qiao}}, \bibinfo {author}
  {\bibfnamefont {L.}~\bibnamefont {Niu}}, \bibinfo {author} {\bibfnamefont
  {C.}~\bibnamefont {Zhao}}, \bibinfo {author} {\bibfnamefont {G.}~\bibnamefont
  {Xue}}, \bibinfo {author} {\bibfnamefont {K.}~\bibnamefont {Bi}}, \bibinfo
  {author} {\bibfnamefont {L.}~\bibnamefont {Mei}}, \bibinfo {author}
  {\bibfnamefont {X.}~\bibnamefont {Wang}}, \ and\ \bibinfo {author}
  {\bibfnamefont {X.}~\bibnamefont {Chou}},\ }\href {\doibase
  https://doi.org/10.1109/LED.2021.3118384} {\bibfield  {journal} {\bibinfo
  {journal} {IEEE Electron Device Letters}\ }\textbf {\bibinfo {volume} {42}},\
  \bibinfo {pages} {1841} (\bibinfo {year} {2021})}\BibitemShut {NoStop}%
\bibitem [{\citenamefont {Shur}\ \emph {et~al.}(2015)\citenamefont {Shur},
  \citenamefont {Akhmatkhanov},\ and\ \citenamefont {Baturin}}]{Shur2015}%
  \BibitemOpen
  \bibfield  {author} {\bibinfo {author} {\bibfnamefont {V.~Y.}\ \bibnamefont
  {Shur}}, \bibinfo {author} {\bibfnamefont {A.~R.}\ \bibnamefont
  {Akhmatkhanov}}, \ and\ \bibinfo {author} {\bibfnamefont {I.~S.}\
  \bibnamefont {Baturin}},\ }\href {\doibase https://doi.org/10.1063/1.4928591}
  {\bibfield  {journal} {\bibinfo  {journal} {Applied Physics Reviews}\
  }\textbf {\bibinfo {volume} {2}},\ \bibinfo {pages} {040604} (\bibinfo {year}
  {2015})}\BibitemShut {NoStop}%
\bibitem [{\citenamefont {Werner}\ \emph {et~al.}(2017)\citenamefont {Werner},
  \citenamefont {Herr}, \citenamefont {Buse}, \citenamefont {Sturman},
  \citenamefont {Soergel}, \citenamefont {Razzaghi},\ and\ \citenamefont
  {Breunig}}]{Werner2017}%
  \BibitemOpen
  \bibfield  {author} {\bibinfo {author} {\bibfnamefont {C.}~\bibnamefont
  {Werner}}, \bibinfo {author} {\bibfnamefont {S.~J.}\ \bibnamefont {Herr}},
  \bibinfo {author} {\bibfnamefont {K.}~\bibnamefont {Buse}}, \bibinfo {author}
  {\bibfnamefont {B.}~\bibnamefont {Sturman}}, \bibinfo {author} {\bibfnamefont
  {E.}~\bibnamefont {Soergel}}, \bibinfo {author} {\bibfnamefont
  {C.}~\bibnamefont {Razzaghi}}, \ and\ \bibinfo {author} {\bibfnamefont
  {I.}~\bibnamefont {Breunig}},\ }\href {\doibase
  https://doi.org/10.1038/s41598-017-09703-2} {\bibfield  {journal} {\bibinfo
  {journal} {Science Reports}\ }\textbf {\bibinfo {volume} {7}},\ \bibinfo
  {pages} {9862} (\bibinfo {year} {2017})}\BibitemShut {NoStop}%
\bibitem [{\citenamefont {Godau}\ \emph {et~al.}(2017)\citenamefont {Godau},
  \citenamefont {Kämpfe}, \citenamefont {Thiessen}, \citenamefont {Eng},\ and\
  \citenamefont {Haußmann}}]{Godau2017}%
  \BibitemOpen
  \bibfield  {author} {\bibinfo {author} {\bibfnamefont {C.}~\bibnamefont
  {Godau}}, \bibinfo {author} {\bibfnamefont {T.}~\bibnamefont {Kämpfe}},
  \bibinfo {author} {\bibfnamefont {A.}~\bibnamefont {Thiessen}}, \bibinfo
  {author} {\bibfnamefont {L.~M.}\ \bibnamefont {Eng}}, \ and\ \bibinfo
  {author} {\bibfnamefont {A.}~\bibnamefont {Haußmann}},\ }\href {\doibase
  https://doi.org/10.1021/acsnano.7b01199} {\bibfield  {journal} {\bibinfo
  {journal} {ACS Nano}\ }\textbf {\bibinfo {volume} {11}},\ \bibinfo {pages}
  {4816} (\bibinfo {year} {2017})}\BibitemShut {NoStop}%
\bibitem [{\citenamefont {Zhang}\ \emph {et~al.}(2022)\citenamefont {Zhang},
  \citenamefont {Qian}, \citenamefont {Jiao}, \citenamefont {Wang},
  \citenamefont {Gao}, \citenamefont {Bo}, \citenamefont {Xu},\ and\
  \citenamefont {Zhang}}]{Zhang2022}%
  \BibitemOpen
  \bibfield  {author} {\bibinfo {author} {\bibfnamefont {Y.}~\bibnamefont
  {Zhang}}, \bibinfo {author} {\bibfnamefont {Y.}~\bibnamefont {Qian}},
  \bibinfo {author} {\bibfnamefont {Y.}~\bibnamefont {Jiao}}, \bibinfo {author}
  {\bibfnamefont {X.}~\bibnamefont {Wang}}, \bibinfo {author} {\bibfnamefont
  {F.}~\bibnamefont {Gao}}, \bibinfo {author} {\bibfnamefont {F.}~\bibnamefont
  {Bo}}, \bibinfo {author} {\bibfnamefont {J.}~\bibnamefont {Xu}}, \ and\
  \bibinfo {author} {\bibfnamefont {G.}~\bibnamefont {Zhang}},\ }\href
  {\doibase https://doi.org/10.1063/5.0101067} {\bibfield  {journal} {\bibinfo
  {journal} {Journal Applied Physics}\ }\textbf {\bibinfo {volume} {132}},\
  \bibinfo {pages} {044102} (\bibinfo {year} {2022})}\BibitemShut {NoStop}%
\bibitem [{\citenamefont {Zhang}\ \emph {et~al.}(2021)\citenamefont {Zhang},
  \citenamefont {Wang}, \citenamefont {Lian}, \citenamefont {Jiang}, ,\ and\
  \citenamefont {Jiang}}]{Zhang2021}%
  \BibitemOpen
  \bibfield  {author} {\bibinfo {author} {\bibfnamefont {W.}~\bibnamefont
  {Zhang}}, \bibinfo {author} {\bibfnamefont {C.}~\bibnamefont {Wang}},
  \bibinfo {author} {\bibfnamefont {J.}~\bibnamefont {Lian}}, \bibinfo {author}
  {\bibfnamefont {J.}~\bibnamefont {Jiang}}, , \ and\ \bibinfo {author}
  {\bibfnamefont {A.-Q.}\ \bibnamefont {Jiang}},\ }\href {\doibase
  https://doi.org/10.1088/0256-307X/38/1/017701} {\bibfield  {journal}
  {\bibinfo  {journal} {Chinese Physics Letters}\ }\textbf {\bibinfo {volume}
  {38}},\ \bibinfo {pages} {017701} (\bibinfo {year} {2021})}\BibitemShut
  {NoStop}%
\bibitem [{\citenamefont {Sharma}\ and\ \citenamefont
  {Seidel}(2023)}]{Sharma2023}%
  \BibitemOpen
  \bibfield  {author} {\bibinfo {author} {\bibfnamefont {P.}~\bibnamefont
  {Sharma}}\ and\ \bibinfo {author} {\bibfnamefont {J.}~\bibnamefont
  {Seidel}},\ }\href {\doibase https://doi.org/10.1088/2634-4386/accfbb}
  {\bibfield  {journal} {\bibinfo  {journal} {Neuromorphic Computing and
  Engineering}\ }\textbf {\bibinfo {volume} {3}},\ \bibinfo {pages} {022001}
  (\bibinfo {year} {2023})}\BibitemShut {NoStop}%
\bibitem [{\citenamefont {Chai}\ \emph {et~al.}(2021)\citenamefont {Chai},
  \citenamefont {Lian}, \citenamefont {Wang}, \citenamefont {Hu}, \citenamefont
  {Sun}, \citenamefont {Jiang},\ and\ \citenamefont {Jiang}}]{Chai2021}%
  \BibitemOpen
  \bibfield  {author} {\bibinfo {author} {\bibfnamefont {X.}~\bibnamefont
  {Chai}}, \bibinfo {author} {\bibfnamefont {J.}~\bibnamefont {Lian}}, \bibinfo
  {author} {\bibfnamefont {C.}~\bibnamefont {Wang}}, \bibinfo {author}
  {\bibfnamefont {X.}~\bibnamefont {Hu}}, \bibinfo {author} {\bibfnamefont
  {J.}~\bibnamefont {Sun}}, \bibinfo {author} {\bibfnamefont {J.}~\bibnamefont
  {Jiang}}, \ and\ \bibinfo {author} {\bibfnamefont {A.}~\bibnamefont
  {Jiang}},\ }\href {\doibase https://doi.org/10.1016/j.jallcom.2021.159837}
  {\bibfield  {journal} {\bibinfo  {journal} {Journal of Alloys and Compounds}\
  }\textbf {\bibinfo {volume} {873}},\ \bibinfo {pages} {159837} (\bibinfo
  {year} {2021})}\BibitemShut {NoStop}%
\bibitem [{\citenamefont {Suna}\ \emph {et~al.}(2022)\citenamefont {Suna},
  \citenamefont {Baxter}, \citenamefont {McConville}, \citenamefont {Kumar},
  \citenamefont {McQuaid},\ and\ \citenamefont {Gregg}}]{Suna2022}%
  \BibitemOpen
  \bibfield  {author} {\bibinfo {author} {\bibfnamefont {A.}~\bibnamefont
  {Suna}}, \bibinfo {author} {\bibfnamefont {O.}~\bibnamefont {Baxter}},
  \bibinfo {author} {\bibfnamefont {J.}~\bibnamefont {McConville}}, \bibinfo
  {author} {\bibfnamefont {A.}~\bibnamefont {Kumar}}, \bibinfo {author}
  {\bibfnamefont {R.}~\bibnamefont {McQuaid}}, \ and\ \bibinfo {author}
  {\bibfnamefont {J.~M.}\ \bibnamefont {Gregg}},\ }\href {\doibase
  https://doi.org/10.1063/5.0124390} {\bibfield  {journal} {\bibinfo  {journal}
  {Applied Physics Letters}\ }\textbf {\bibinfo {volume} {121}},\ \bibinfo
  {pages} {222902} (\bibinfo {year} {2022})}\BibitemShut {NoStop}%
\bibitem [{\citenamefont {Turner}\ \emph {et~al.}(1994)\citenamefont {Turner},
  \citenamefont {Fuierer},\ and\ \citenamefont {Newnham}}]{Turner1994}%
  \BibitemOpen
  \bibfield  {author} {\bibinfo {author} {\bibfnamefont {R.}~\bibnamefont
  {Turner}}, \bibinfo {author} {\bibfnamefont {P.}~\bibnamefont {Fuierer}}, \
  and\ \bibinfo {author} {\bibfnamefont {R.}~\bibnamefont {Newnham}},\ }\href
  {\doibase https://doi.org/10.1016/0003-682X(94)90091-4} {\bibfield  {journal}
  {\bibinfo  {journal} {Applied Acoustics}\ }\textbf {\bibinfo {volume} {41}},\
  \bibinfo {pages} {299} (\bibinfo {year} {1994})}\BibitemShut {NoStop}%
\bibitem [{\citenamefont {Zahn}\ \emph {et~al.}(2024)\citenamefont {Zahn},
  \citenamefont {Beyreuther}, \citenamefont {Kiseleva}, \citenamefont {Lotfy},
  \citenamefont {McCluskey}, \citenamefont {Maguire}, \citenamefont {A.Suna},
  \citenamefont {Rüsing}, \citenamefont {Gregg},\ and\ \citenamefont
  {Eng}}]{Zahn2024}%
  \BibitemOpen
  \bibfield  {author} {\bibinfo {author} {\bibfnamefont {M.}~\bibnamefont
  {Zahn}}, \bibinfo {author} {\bibfnamefont {E.}~\bibnamefont {Beyreuther}},
  \bibinfo {author} {\bibfnamefont {I.}~\bibnamefont {Kiseleva}}, \bibinfo
  {author} {\bibfnamefont {A.}~\bibnamefont {Lotfy}}, \bibinfo {author}
  {\bibfnamefont {C.}~\bibnamefont {McCluskey}}, \bibinfo {author}
  {\bibfnamefont {J.}~\bibnamefont {Maguire}}, \bibinfo {author} {\bibnamefont
  {A.Suna}}, \bibinfo {author} {\bibfnamefont {M.}~\bibnamefont {Rüsing}},
  \bibinfo {author} {\bibfnamefont {J.}~\bibnamefont {Gregg}}, \ and\ \bibinfo
  {author} {\bibfnamefont {L.}~\bibnamefont {Eng}},\ }\href {\doibase
  https://doi.org/10.1103/PhysRevApplied.21.024007} {\bibfield  {journal}
  {\bibinfo  {journal} {Physical Review Applied}\ }\textbf {\bibinfo {volume}
  {21}},\ \bibinfo {pages} {024007} (\bibinfo {year} {2024})}\BibitemShut
  {NoStop}%
\bibitem [{\citenamefont {Beccard}\ \emph {et~al.}(2023)\citenamefont
  {Beccard}, \citenamefont {Beyreuther}, \citenamefont {Kirbus}, \citenamefont
  {Seddon}, \citenamefont {Rüsing},\ and\ \citenamefont {Eng}}]{Beccard2023}%
  \BibitemOpen
  \bibfield  {author} {\bibinfo {author} {\bibfnamefont {H.}~\bibnamefont
  {Beccard}}, \bibinfo {author} {\bibfnamefont {E.}~\bibnamefont {Beyreuther}},
  \bibinfo {author} {\bibfnamefont {B.}~\bibnamefont {Kirbus}}, \bibinfo
  {author} {\bibfnamefont {S.~D.}\ \bibnamefont {Seddon}}, \bibinfo {author}
  {\bibfnamefont {M.}~\bibnamefont {Rüsing}}, \ and\ \bibinfo {author}
  {\bibfnamefont {L.~M.}\ \bibnamefont {Eng}},\ }\href {\doibase
  10.1103/PhysRevApplied.20.064043} {\bibfield  {journal} {\bibinfo  {journal}
  {Physical Review Applied}\ }\textbf {\bibinfo {volume} {20}},\ \bibinfo
  {pages} {064043} (\bibinfo {year} {2023})}\BibitemShut {NoStop}%
\bibitem [{\citenamefont {Nataf}\ \emph {et~al.}(2020)\citenamefont {Nataf},
  \citenamefont {Guennou}, \citenamefont {Gregg}, \citenamefont {Meier},
  \citenamefont {Hlinka}, \citenamefont {Salje},\ and\ \citenamefont
  {Kreisel}}]{nataf20}%
  \BibitemOpen
  \bibfield  {author} {\bibinfo {author} {\bibfnamefont {G.}~\bibnamefont
  {Nataf}}, \bibinfo {author} {\bibfnamefont {M.}~\bibnamefont {Guennou}},
  \bibinfo {author} {\bibfnamefont {J.}~\bibnamefont {Gregg}}, \bibinfo
  {author} {\bibfnamefont {D.}~\bibnamefont {Meier}}, \bibinfo {author}
  {\bibfnamefont {J.}~\bibnamefont {Hlinka}}, \bibinfo {author} {\bibfnamefont
  {E.~K.~H.}\ \bibnamefont {Salje}}, \ and\ \bibinfo {author} {\bibfnamefont
  {J.}~\bibnamefont {Kreisel}},\ }\href {\doibase 10.1038/s42254-020-0235-z}
  {\bibfield  {journal} {\bibinfo  {journal} {Nature Reviews Physics}\ }\textbf
  {\bibinfo {volume} {2}},\ \bibinfo {pages} {634} (\bibinfo {year}
  {2020})}\BibitemShut {NoStop}%
\bibitem [{\citenamefont {Bednyakov}\ \emph {et~al.}(2018)\citenamefont
  {Bednyakov}, \citenamefont {Sturman}, \citenamefont {Sluka}, \citenamefont
  {Tagantsev},\ and\ \citenamefont {Yudin}}]{bed18}%
  \BibitemOpen
  \bibfield  {author} {\bibinfo {author} {\bibfnamefont {P.~S.}\ \bibnamefont
  {Bednyakov}}, \bibinfo {author} {\bibfnamefont {B.~I.}\ \bibnamefont
  {Sturman}}, \bibinfo {author} {\bibfnamefont {T.}~\bibnamefont {Sluka}},
  \bibinfo {author} {\bibfnamefont {A.~K.}\ \bibnamefont {Tagantsev}}, \ and\
  \bibinfo {author} {\bibfnamefont {P.~V.}\ \bibnamefont {Yudin}},\ }\href
  {\doibase 10.1038/s41524-018-0121-8} {\bibfield  {journal} {\bibinfo
  {journal} {npj Computational Materials}\ }\textbf {\bibinfo {volume} {4}},\
  \bibinfo {pages} {65} (\bibinfo {year} {2018})}\BibitemShut {NoStop}%
\bibitem [{\citenamefont {Beccard}\ \emph {et~al.}(2022)\citenamefont
  {Beccard}, \citenamefont {Kirbus}, \citenamefont {Beyreuther}, \citenamefont
  {R\"using}, \citenamefont {Bednyakov}, \citenamefont {Hlinka},\ and\
  \citenamefont {Eng}}]{bec22}%
  \BibitemOpen
  \bibfield  {author} {\bibinfo {author} {\bibfnamefont {H.}~\bibnamefont
  {Beccard}}, \bibinfo {author} {\bibfnamefont {B.}~\bibnamefont {Kirbus}},
  \bibinfo {author} {\bibfnamefont {E.}~\bibnamefont {Beyreuther}}, \bibinfo
  {author} {\bibfnamefont {M.}~\bibnamefont {R\"using}}, \bibinfo {author}
  {\bibfnamefont {P.}~\bibnamefont {Bednyakov}}, \bibinfo {author}
  {\bibfnamefont {J.}~\bibnamefont {Hlinka}}, \ and\ \bibinfo {author}
  {\bibfnamefont {L.~M.}\ \bibnamefont {Eng}},\ }\href {\doibase
  10.1021/acsanm.2c01919} {\bibfield  {journal} {\bibinfo  {journal} {ACS
  Applied Nano Materials}\ }\textbf {\bibinfo {volume} {5}},\ \bibinfo {pages}
  {8717} (\bibinfo {year} {2022})}\BibitemShut {NoStop}%
\bibitem [{\citenamefont {Xiao}\ \emph {et~al.}(2018)\citenamefont {Xiao},
  \citenamefont {K\"ampfe}, \citenamefont {Jin}, \citenamefont {Hau\ss{}mann},
  \citenamefont {Lu},\ and\ \citenamefont {Eng}}]{xia18}%
  \BibitemOpen
  \bibfield  {author} {\bibinfo {author} {\bibfnamefont {S.~Y.}\ \bibnamefont
  {Xiao}}, \bibinfo {author} {\bibfnamefont {T.}~\bibnamefont {K\"ampfe}},
  \bibinfo {author} {\bibfnamefont {Y.~M.}\ \bibnamefont {Jin}}, \bibinfo
  {author} {\bibfnamefont {A.}~\bibnamefont {Hau\ss{}mann}}, \bibinfo {author}
  {\bibfnamefont {X.~M.}\ \bibnamefont {Lu}}, \ and\ \bibinfo {author}
  {\bibfnamefont {L.~M.}\ \bibnamefont {Eng}},\ }\href {\doibase
  10.1103/PhysRevApplied.10.034002} {\bibfield  {journal} {\bibinfo  {journal}
  {Physical Review Applied}\ }\textbf {\bibinfo {volume} {10}},\ \bibinfo
  {pages} {034002} (\bibinfo {year} {2018})}\BibitemShut {NoStop}%
\bibitem [{\citenamefont {Haußmann}\ \emph {et~al.}(2009)\citenamefont
  {Haußmann}, \citenamefont {Milde}, \citenamefont {Erler},\ and\
  \citenamefont {Eng}}]{Haussmann2009}%
  \BibitemOpen
  \bibfield  {author} {\bibinfo {author} {\bibfnamefont {A.}~\bibnamefont
  {Haußmann}}, \bibinfo {author} {\bibfnamefont {P.}~\bibnamefont {Milde}},
  \bibinfo {author} {\bibfnamefont {C.}~\bibnamefont {Erler}}, \ and\ \bibinfo
  {author} {\bibfnamefont {L.}~\bibnamefont {Eng}},\ }\href {\doibase
  https://doi.org/10.1103/PhysRevApplied.21.024007} {\bibfield  {journal}
  {\bibinfo  {journal} {Nano Letters}\ }\textbf {\bibinfo {volume} {9}},\
  \bibinfo {pages} {763} (\bibinfo {year} {2009})}\BibitemShut {NoStop}%
\bibitem [{\citenamefont {Yakhnevych}\ \emph {et~al.}(2024)\citenamefont
  {Yakhnevych}, \citenamefont {Azzouzi}, \citenamefont {Bernhardt},
  \citenamefont {Kofahl}, \citenamefont {Suhak}, \citenamefont {Sanna},
  \citenamefont {Becker}, \citenamefont {Schmidt}, \citenamefont {Ganschow},\
  and\ \citenamefont {Fritze}}]{Yakhnevych2024}%
  \BibitemOpen
  \bibfield  {author} {\bibinfo {author} {\bibfnamefont {U.}~\bibnamefont
  {Yakhnevych}}, \bibinfo {author} {\bibfnamefont {F.~E.}\ \bibnamefont
  {Azzouzi}}, \bibinfo {author} {\bibfnamefont {F.}~\bibnamefont {Bernhardt}},
  \bibinfo {author} {\bibfnamefont {C.}~\bibnamefont {Kofahl}}, \bibinfo
  {author} {\bibfnamefont {Y.}~\bibnamefont {Suhak}}, \bibinfo {author}
  {\bibfnamefont {S.}~\bibnamefont {Sanna}}, \bibinfo {author} {\bibfnamefont
  {K.-D.}\ \bibnamefont {Becker}}, \bibinfo {author} {\bibfnamefont
  {H.}~\bibnamefont {Schmidt}}, \bibinfo {author} {\bibfnamefont
  {S.}~\bibnamefont {Ganschow}}, \ and\ \bibinfo {author} {\bibfnamefont
  {H.}~\bibnamefont {Fritze}},\ }\href {\doibase
  https://doi.org/10.1016/j.ssi.2024.116487} {\bibfield  {journal} {\bibinfo
  {journal} {Solid State Ionics}\ }\textbf {\bibinfo {volume} {407}},\ \bibinfo
  {pages} {116487} (\bibinfo {year} {2024})}\BibitemShut {NoStop}%
\bibitem [{\citenamefont {Kofahl}\ \emph {et~al.}(2024)\citenamefont {Kofahl},
  \citenamefont {Dörrer}, \citenamefont {Wulfmeier}, \citenamefont {Fritze},
  \citenamefont {Ganschow},\ and\ \citenamefont {Schmidt}}]{Kofahl2024}%
  \BibitemOpen
  \bibfield  {author} {\bibinfo {author} {\bibfnamefont {C.}~\bibnamefont
  {Kofahl}}, \bibinfo {author} {\bibfnamefont {L.}~\bibnamefont {Dörrer}},
  \bibinfo {author} {\bibfnamefont {H.}~\bibnamefont {Wulfmeier}}, \bibinfo
  {author} {\bibfnamefont {H.}~\bibnamefont {Fritze}}, \bibinfo {author}
  {\bibfnamefont {S.}~\bibnamefont {Ganschow}}, \ and\ \bibinfo {author}
  {\bibfnamefont {H.}~\bibnamefont {Schmidt}},\ }\href {\doibase
  https://doi.org/10.1021/acs.chemmater.3c02984} {\bibfield  {journal}
  {\bibinfo  {journal} {Chemistry of Materials}\ }\textbf {\bibinfo {volume}
  {36}},\ \bibinfo {pages} {1639} (\bibinfo {year} {2024})}\BibitemShut
  {NoStop}%
\bibitem [{\citenamefont {Singh}\ \emph {et~al.}(2022)\citenamefont {Singh},
  \citenamefont {Beccard}, \citenamefont {Amber}, \citenamefont {Ratzenberger},
  \citenamefont {Hicks}, \citenamefont {R\"using},\ and\ \citenamefont
  {Eng}}]{sin22}%
  \BibitemOpen
  \bibfield  {author} {\bibinfo {author} {\bibfnamefont {E.}~\bibnamefont
  {Singh}}, \bibinfo {author} {\bibfnamefont {H.}~\bibnamefont {Beccard}},
  \bibinfo {author} {\bibfnamefont {Z.~H.}\ \bibnamefont {Amber}}, \bibinfo
  {author} {\bibfnamefont {J.}~\bibnamefont {Ratzenberger}}, \bibinfo {author}
  {\bibfnamefont {C.~W.}\ \bibnamefont {Hicks}}, \bibinfo {author}
  {\bibfnamefont {M.}~\bibnamefont {R\"using}}, \ and\ \bibinfo {author}
  {\bibfnamefont {L.~M.}\ \bibnamefont {Eng}},\ }\href {\doibase
  10.1103/PhysRevB.106.144103} {\bibfield  {journal} {\bibinfo  {journal}
  {Physical Review B}\ }\textbf {\bibinfo {volume} {106}},\ \bibinfo {pages}
  {144103} (\bibinfo {year} {2022})}\BibitemShut {NoStop}%
\bibitem [{\citenamefont {Reichenbach}\ \emph {et~al.}(2018)\citenamefont
  {Reichenbach}, \citenamefont {Kämpfe}, \citenamefont {Haußmann},
  \citenamefont {Thiessen}, \citenamefont {Woike}, \citenamefont {Steudtner},
  \citenamefont {Kocsor}, \citenamefont {Szaller}, \citenamefont {Kovács},\
  and\ \citenamefont {Eng}}]{Reichenbach2018}%
  \BibitemOpen
  \bibfield  {author} {\bibinfo {author} {\bibfnamefont {P.}~\bibnamefont
  {Reichenbach}}, \bibinfo {author} {\bibfnamefont {T.}~\bibnamefont
  {Kämpfe}}, \bibinfo {author} {\bibfnamefont {A.}~\bibnamefont {Haußmann}},
  \bibinfo {author} {\bibfnamefont {A.}~\bibnamefont {Thiessen}}, \bibinfo
  {author} {\bibfnamefont {T.}~\bibnamefont {Woike}}, \bibinfo {author}
  {\bibfnamefont {R.}~\bibnamefont {Steudtner}}, \bibinfo {author}
  {\bibfnamefont {L.}~\bibnamefont {Kocsor}}, \bibinfo {author} {\bibfnamefont
  {Z.}~\bibnamefont {Szaller}}, \bibinfo {author} {\bibfnamefont
  {L.}~\bibnamefont {Kovács}}, \ and\ \bibinfo {author} {\bibfnamefont
  {L.~M.}\ \bibnamefont {Eng}},\ }\href {\doibase
  https://doi.org/10.3390/cryst8050214} {\bibfield  {journal} {\bibinfo
  {journal} {Crystals}\ }\textbf {\bibinfo {volume} {8}},\ \bibinfo {pages}
  {214} (\bibinfo {year} {2018})}\BibitemShut {NoStop}%
\bibitem [{\citenamefont {Koppitz}\ \emph {et~al.}(1987)\citenamefont
  {Koppitz}, \citenamefont {Schirmer},\ and\ \citenamefont
  {Kuznetsov}}]{Koppitz1987}%
  \BibitemOpen
  \bibfield  {author} {\bibinfo {author} {\bibfnamefont {J.}~\bibnamefont
  {Koppitz}}, \bibinfo {author} {\bibfnamefont {O.}~\bibnamefont {Schirmer}}, \
  and\ \bibinfo {author} {\bibfnamefont {A.}~\bibnamefont {Kuznetsov}},\ }\href
  {\doibase https://doi.org/10.1209/0295-5075/4/9/017} {\bibfield  {journal}
  {\bibinfo  {journal} {Europhysics Letters}\ }\textbf {\bibinfo {volume}
  {4}},\ \bibinfo {pages} {1055–1059} (\bibinfo {year} {1987})}\BibitemShut
  {NoStop}%
\bibitem [{\citenamefont {Schirmer}\ \emph {et~al.}(2009)\citenamefont
  {Schirmer}, \citenamefont {Imlau}, \citenamefont {Merschjann},\ and\
  \citenamefont {Schoke}}]{Schirmer2009}%
  \BibitemOpen
  \bibfield  {author} {\bibinfo {author} {\bibfnamefont {O.}~\bibnamefont
  {Schirmer}}, \bibinfo {author} {\bibfnamefont {M.}~\bibnamefont {Imlau}},
  \bibinfo {author} {\bibfnamefont {C.}~\bibnamefont {Merschjann}}, \ and\
  \bibinfo {author} {\bibfnamefont {B.}~\bibnamefont {Schoke}},\ }\href
  {\doibase 10.1088/0953-8984/21/12/123201} {\bibfield  {journal} {\bibinfo
  {journal} {Journal of Physics: Condensed Matter}\ }\textbf {\bibinfo {volume}
  {21}},\ \bibinfo {pages} {123201} (\bibinfo {year} {2009})}\BibitemShut
  {NoStop}%
\bibitem [{\citenamefont {Kresse}\ and\ \citenamefont
  {Furthm{\"{u}}ller}(1996{\natexlab{a}})}]{Kresse1996}%
  \BibitemOpen
  \bibfield  {author} {\bibinfo {author} {\bibfnamefont {G.}~\bibnamefont
  {Kresse}}\ and\ \bibinfo {author} {\bibfnamefont {J.}~\bibnamefont
  {Furthm{\"{u}}ller}},\ }\href {\doibase 10.1103/PhysRevB.54.11169} {\bibfield
   {journal} {\bibinfo  {journal} {Physical Review B}\ }\textbf {\bibinfo
  {volume} {54}},\ \bibinfo {pages} {11169} (\bibinfo {year}
  {1996}{\natexlab{a}})}\BibitemShut {NoStop}%
\bibitem [{\citenamefont {Kresse}\ and\ \citenamefont
  {Furthm{\"{u}}ller}(1996{\natexlab{b}})}]{Kresse1996_2}%
  \BibitemOpen
  \bibfield  {author} {\bibinfo {author} {\bibfnamefont {G.}~\bibnamefont
  {Kresse}}\ and\ \bibinfo {author} {\bibfnamefont {J.}~\bibnamefont
  {Furthm{\"{u}}ller}},\ }\href {\doibase 10.1016/0927-0256(96)00008-0}
  {\bibfield  {journal} {\bibinfo  {journal} {Computational Materials Science}\
  }\textbf {\bibinfo {volume} {6}},\ \bibinfo {pages} {15} (\bibinfo {year}
  {1996}{\natexlab{b}})}\BibitemShut {NoStop}%
\bibitem [{\citenamefont {Perdew}\ \emph {et~al.}(1996)\citenamefont {Perdew},
  \citenamefont {Burke},\ and\ \citenamefont {Ernzerhof}}]{Perdew1996}%
  \BibitemOpen
  \bibfield  {author} {\bibinfo {author} {\bibfnamefont {J.~P.}\ \bibnamefont
  {Perdew}}, \bibinfo {author} {\bibfnamefont {K.}~\bibnamefont {Burke}}, \
  and\ \bibinfo {author} {\bibfnamefont {M.}~\bibnamefont {Ernzerhof}},\ }\href
  {\doibase 10.1103/PhysRevLett.77.3865} {\bibfield  {journal} {\bibinfo
  {journal} {Physical Review Letters}\ }\textbf {\bibinfo {volume} {77}},\
  \bibinfo {pages} {3865} (\bibinfo {year} {1996})}\BibitemShut {NoStop}%
\bibitem [{\citenamefont {Perdew}\ \emph {et~al.}(2008)\citenamefont {Perdew},
  \citenamefont {Ruzsinszky}, \citenamefont {Csonka}, \citenamefont {Vydrov},
  \citenamefont {Scuseria}, \citenamefont {Constantin}, \citenamefont {Zhou},\
  and\ \citenamefont {Burke}}]{Perdew2008}%
  \BibitemOpen
  \bibfield  {author} {\bibinfo {author} {\bibfnamefont {J.~P.}\ \bibnamefont
  {Perdew}}, \bibinfo {author} {\bibfnamefont {A.}~\bibnamefont {Ruzsinszky}},
  \bibinfo {author} {\bibfnamefont {G.~I.}\ \bibnamefont {Csonka}}, \bibinfo
  {author} {\bibfnamefont {O.~A.}\ \bibnamefont {Vydrov}}, \bibinfo {author}
  {\bibfnamefont {G.~E.}\ \bibnamefont {Scuseria}}, \bibinfo {author}
  {\bibfnamefont {L.~A.}\ \bibnamefont {Constantin}}, \bibinfo {author}
  {\bibfnamefont {X.}~\bibnamefont {Zhou}}, \ and\ \bibinfo {author}
  {\bibfnamefont {K.}~\bibnamefont {Burke}},\ }\href {\doibase
  10.1103/PhysRevLett.100.136406} {\bibfield  {journal} {\bibinfo  {journal}
  {Physical Review Letters}\ }\textbf {\bibinfo {volume} {100}},\ \bibinfo
  {pages} {136406} (\bibinfo {year} {2008})}\BibitemShut {NoStop}%
\bibitem [{\citenamefont {Bl\"ochl}(1994)}]{Bloechl94}%
  \BibitemOpen
  \bibfield  {author} {\bibinfo {author} {\bibfnamefont {P.~E.}\ \bibnamefont
  {Bl\"ochl}},\ }\href {\doibase 10.1103/PhysRevB.50.17953} {\bibfield
  {journal} {\bibinfo  {journal} {Physical Review B}\ }\textbf {\bibinfo
  {volume} {50}},\ \bibinfo {pages} {17953} (\bibinfo {year}
  {1994})}\BibitemShut {NoStop}%
\bibitem [{\citenamefont {Gonnissen}\ \emph {et~al.}(2016)\citenamefont
  {Gonnissen}, \citenamefont {Batuk}, \citenamefont {Nataf}, \citenamefont
  {Jones}, \citenamefont {Abakumov}, \citenamefont {Van~Aert}, \citenamefont
  {Schryvers},\ and\ \citenamefont {Salje}}]{Gonnissen2016}%
  \BibitemOpen
  \bibfield  {author} {\bibinfo {author} {\bibfnamefont {J.}~\bibnamefont
  {Gonnissen}}, \bibinfo {author} {\bibfnamefont {D.}~\bibnamefont {Batuk}},
  \bibinfo {author} {\bibfnamefont {G.~F.}\ \bibnamefont {Nataf}}, \bibinfo
  {author} {\bibfnamefont {L.}~\bibnamefont {Jones}}, \bibinfo {author}
  {\bibfnamefont {A.~M.}\ \bibnamefont {Abakumov}}, \bibinfo {author}
  {\bibfnamefont {S.}~\bibnamefont {Van~Aert}}, \bibinfo {author}
  {\bibfnamefont {D.}~\bibnamefont {Schryvers}}, \ and\ \bibinfo {author}
  {\bibfnamefont {E.~K.~H.}\ \bibnamefont {Salje}},\ }\href {\doibase
  10.1002/adfm.201603489} {\bibfield  {journal} {\bibinfo  {journal} {Advanced
  Functional Materials}\ }\textbf {\bibinfo {volume} {26}},\ \bibinfo {pages}
  {7599} (\bibinfo {year} {2016})}\BibitemShut {NoStop}%
\bibitem [{\citenamefont {Verhoff}\ \emph {et~al.}()\citenamefont {Verhoff},
  \citenamefont {R\"using}, \citenamefont {Eng},\ and\ \citenamefont
  {Sanna}}]{Verhoff2024}%
  \BibitemOpen
  \bibfield  {author} {\bibinfo {author} {\bibfnamefont {L.~M.}\ \bibnamefont
  {Verhoff}}, \bibinfo {author} {\bibfnamefont {M.}~\bibnamefont {R\"using}},
  \bibinfo {author} {\bibfnamefont {L.}~\bibnamefont {Eng}}, \ and\ \bibinfo
  {author} {\bibfnamefont {S.}~\bibnamefont {Sanna}},\ }\href@noop {} {\bibinfo
   {journal} {in preparation}\ }\BibitemShut {NoStop}%
\bibitem [{\citenamefont {Riefer}\ \emph {et~al.}(2016)\citenamefont {Riefer},
  \citenamefont {Friedrich}, \citenamefont {Sanna}, \citenamefont {Gerstmann},
  \citenamefont {Schindlmayr},\ and\ \citenamefont {Schmidt}}]{Riefer16}%
  \BibitemOpen
\bibfield  {journal} {  }\bibfield  {author} {\bibinfo {author} {\bibfnamefont
  {A.}~\bibnamefont {Riefer}}, \bibinfo {author} {\bibfnamefont
  {M.}~\bibnamefont {Friedrich}}, \bibinfo {author} {\bibfnamefont
  {S.}~\bibnamefont {Sanna}}, \bibinfo {author} {\bibfnamefont
  {U.}~\bibnamefont {Gerstmann}}, \bibinfo {author} {\bibfnamefont
  {A.}~\bibnamefont {Schindlmayr}}, \ and\ \bibinfo {author} {\bibfnamefont
  {W.~G.}\ \bibnamefont {Schmidt}},\ }\href {\doibase
  10.1103/PhysRevB.93.075205} {\bibfield  {journal} {\bibinfo  {journal} {Phys.
  Rev. B}\ }\textbf {\bibinfo {volume} {93}},\ \bibinfo {pages} {075205}
  (\bibinfo {year} {2016})}\BibitemShut {NoStop}%
\bibitem [{\citenamefont {Schröder}\ \emph {et~al.}(2012)\citenamefont
  {Schröder}, \citenamefont {Haußmann}, \citenamefont {Thiessen},
  \citenamefont {Soergel}, \citenamefont {Woike},\ and\ \citenamefont
  {Eng}}]{Schröder2012}%
  \BibitemOpen
  \bibfield  {author} {\bibinfo {author} {\bibfnamefont {M.}~\bibnamefont
  {Schröder}}, \bibinfo {author} {\bibfnamefont {A.}~\bibnamefont
  {Haußmann}}, \bibinfo {author} {\bibfnamefont {A.}~\bibnamefont {Thiessen}},
  \bibinfo {author} {\bibfnamefont {E.}~\bibnamefont {Soergel}}, \bibinfo
  {author} {\bibfnamefont {T.}~\bibnamefont {Woike}}, \ and\ \bibinfo {author}
  {\bibfnamefont {L.}~\bibnamefont {Eng}},\ }\href {\doibase
  https://doi.org/10.1038/nmat1067} {\bibfield  {journal} {\bibinfo  {journal}
  {Advanced Functional Materials}\ }\textbf {\bibinfo {volume} {22}},\ \bibinfo
  {pages} {3936} (\bibinfo {year} {2012})}\BibitemShut {NoStop}%
\bibitem [{\citenamefont {Kiseleva}(2023)}]{Kiseleva2023}%
  \BibitemOpen
  \bibfield  {author} {\bibinfo {author} {\bibfnamefont {I.}~\bibnamefont
  {Kiseleva}},\ }\emph {\bibinfo {title} {{Toward Reproducible Domain-Wall
  Conductance in Lithium Niobate Single Crystals}}},\ \href
  {https://nbn-resolving.org/urn:nbn:de:bsz:14-qucosa2-876252} {\bibinfo {type}
  {{Master's thesis}}},\ \bibinfo  {school} {Technische Universität Dresden},
  \bibinfo {address} {Dresden, Germany} (\bibinfo {year} {2023})\BibitemShut
  {NoStop}%
\bibitem [{\citenamefont {Ding}\ \emph {et~al.}(2024)\citenamefont {Ding},
  \citenamefont {Beyreuther}, \citenamefont {Koppitz}, \citenamefont {Kempf},
  \citenamefont {Ren}, \citenamefont {Chen}, \citenamefont {Rüsing},
  \citenamefont {Zheng},\ and\ \citenamefont {Eng}}]{Ding2024}%
  \BibitemOpen
  \bibfield  {author} {\bibinfo {author} {\bibfnamefont {L.}~\bibnamefont
  {Ding}}, \bibinfo {author} {\bibfnamefont {E.}~\bibnamefont {Beyreuther}},
  \bibinfo {author} {\bibfnamefont {B.}~\bibnamefont {Koppitz}}, \bibinfo
  {author} {\bibfnamefont {K.}~\bibnamefont {Kempf}}, \bibinfo {author}
  {\bibfnamefont {J.}~\bibnamefont {Ren}}, \bibinfo {author} {\bibfnamefont
  {W.}~\bibnamefont {Chen}}, \bibinfo {author} {\bibfnamefont {M.}~\bibnamefont
  {Rüsing}}, \bibinfo {author} {\bibfnamefont {Y.}~\bibnamefont {Zheng}}, \
  and\ \bibinfo {author} {\bibfnamefont {L.~M.}\ \bibnamefont {Eng}},\ }\href
  {\doibase 10.48550/arXiv.2402.17508} {\enquote {\bibinfo {title} {Comparative
  study of photo-induced electronic transport along ferroelectric domain walls
  in lithium niobate single crystals},}\ } (\bibinfo {year} {2024}),\ \Eprint
  {http://arxiv.org/abs/2402.17508} {arXiv:2402.17508 [physics.app-ph]}
  \BibitemShut {NoStop}%
\end{thebibliography}%

\setcounter{figure}{0}
\renewcommand{\thefigure}{S\arabic{figure}}

% Встановлення стилю для підсекцій
\titleformat{\section}
{\normalfont\Large\bfseries}{\thesubsection}{5em}{}

\onecolumngrid
\newpage

\setcounter{page}{1}
\renewcommand{\thepage}{S\arabic{page}} 
%\titleformat{\section}{\centering\normalfont\bfseries}{\thesection.}{.3em}{\vspace{.3ex}}

\section*{Supplementary Material}

\vspace{\baselineskip}

\textbf{Supplement S1: Domain images and current-voltage characteristics of all investigated domains.}

\begin{figure}[H]
	\centering
	\begin{subfigure}{0.005\textwidth}
		\centering
		\includegraphics[scale=0.7]{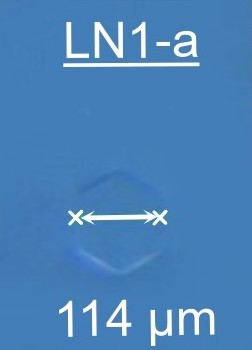}
		\caption{} % Caption for the first subfigure (you can add text here)
		\label{fig:subfig_a1}
	\end{subfigure}
	\hfill
	\begin{subfigure}{0.2\textwidth}
		\centering
		\includegraphics{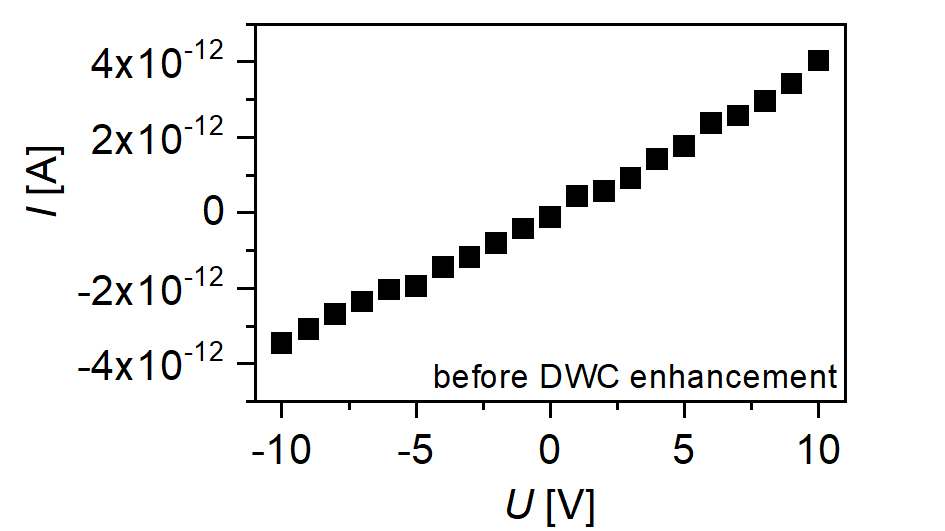}
		\caption{}
		\label{fig:subfig_b1}
	\end{subfigure}
	\hfill
	\begin{subfigure}{0.37\textwidth}
		\centering
		\includegraphics{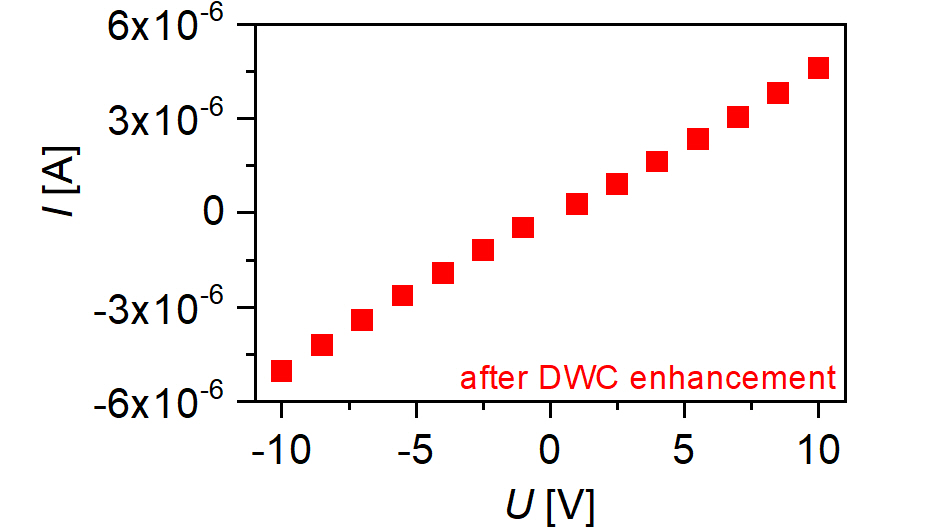}
		\caption{}
		\label{fig:subfig_c1}
	\end{subfigure}
	\caption{\textbf{LN1-a:} (a) Domain image; (b) current-voltage characteristics after poling,
		and (c) after DWC enhancement.} % Overall caption for both subfigures
	\label{fig:1} % Label for the entire figure
\end{figure}

\begin{figure}[H]
	\centering
	\begin{subfigure}{0.005\textwidth}
		\centering
		\includegraphics[scale=0.7]{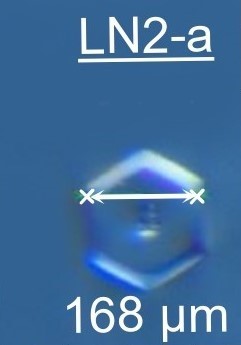}
		\caption{} % Caption for the first subfigure (you can add text here)
		\label{fig:subfig_a2}
	\end{subfigure}
	\hfill
	\begin{subfigure}{0.2\textwidth}
		\centering
		\includegraphics{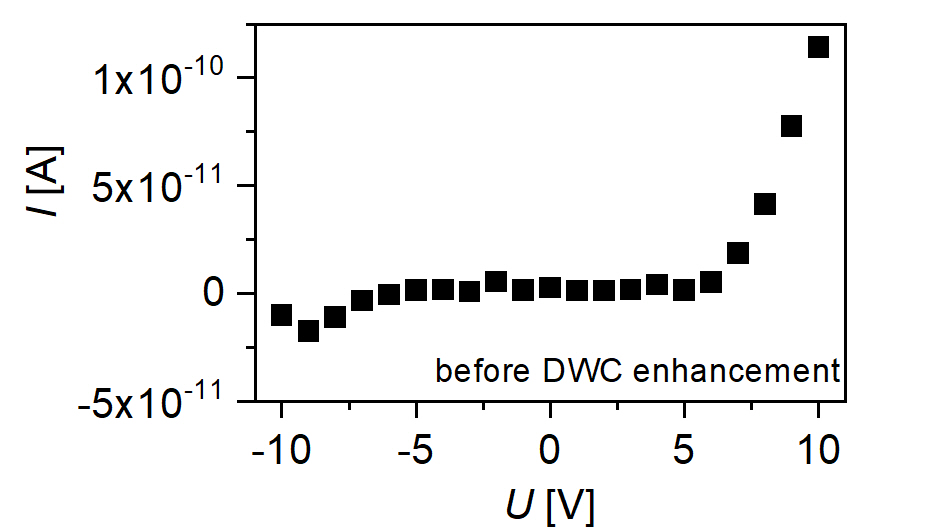}
		\caption{}
		\label{fig:subfig_b2}
	\end{subfigure}
	\hfill
	\begin{subfigure}{0.37\textwidth}
		\centering
		\includegraphics{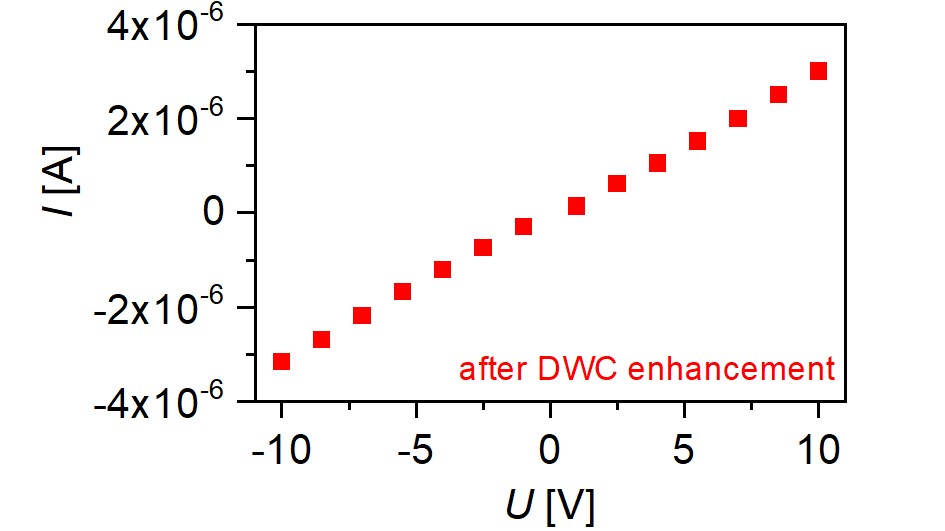}
		\caption{}
		\label{fig:subfig_c2}
	\end{subfigure}
	\caption{\textbf{LN2-a:} (a) Domain image; (b) current-voltage characteristics after poling,
		and (c) after DWC enhancement.} % Overall caption for both subfigures
	\label{fig:2} % Label for the entire figure
\end{figure}

\begin{figure}[H]
	\centering
	\begin{subfigure}{0.005\textwidth}
		\centering
		\includegraphics[scale=0.7]{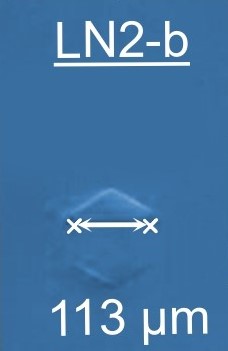}
		\caption{} % Caption for the first subfigure (you can add text here)
		\label{fig:subfig_a3}
	\end{subfigure}
	\hfill
	\begin{subfigure}{0.2\textwidth}
		\centering
		\includegraphics{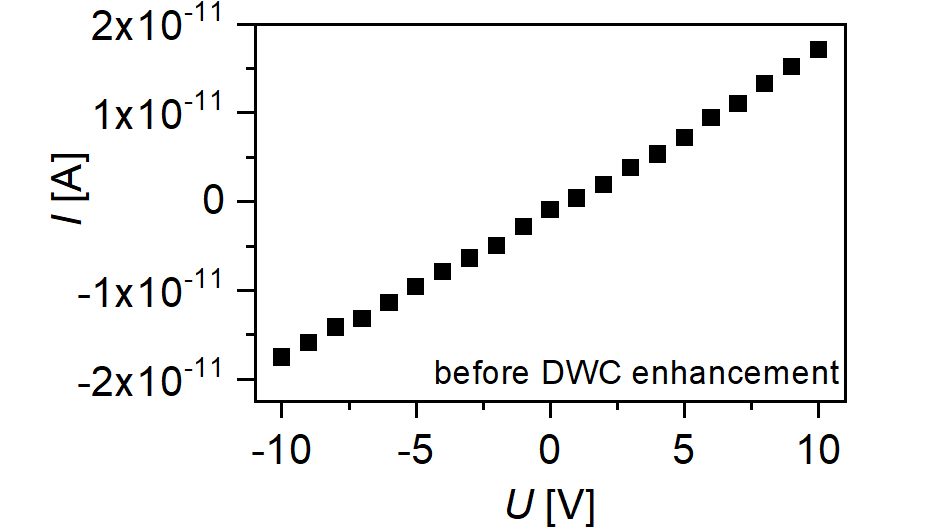}
		\caption{}
		\label{fig:subfig_b3}
	\end{subfigure}
	\hfill
	\begin{subfigure}{0.37\textwidth}
		\centering
		\includegraphics{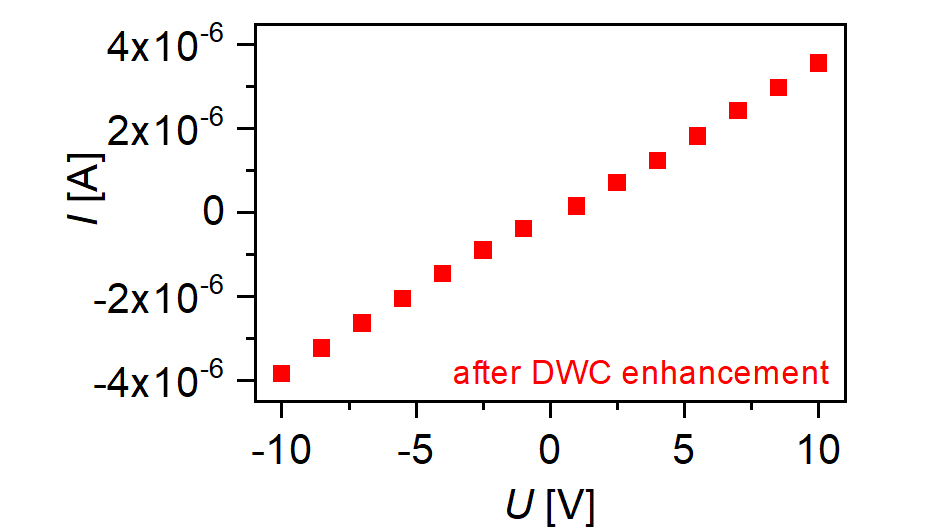}
		\caption{}
		\label{fig:subfig_c3}
	\end{subfigure}
	\caption{\textbf{LN2-b:} (a) Domain image; (b) current-voltage characteristics after poling,
		and (c) after DWC enhancement.} % Overall caption for both subfigures
	\label{fig:3} % Label for the entire figure
\end{figure}

\begin{figure}[H]
	\centering
	\begin{subfigure}{0.005\textwidth}
		\centering
		\includegraphics[scale=0.7]{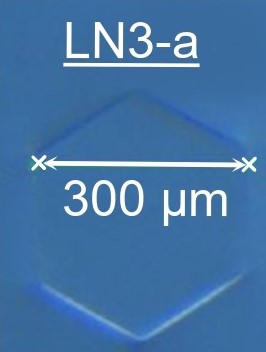}
		\caption{} % Caption for the first subfigure (you can add text here)
		\label{fig:subfig_a}
	\end{subfigure}
	\hfill
	\begin{subfigure}{0.2\textwidth}
		\centering
		\includegraphics{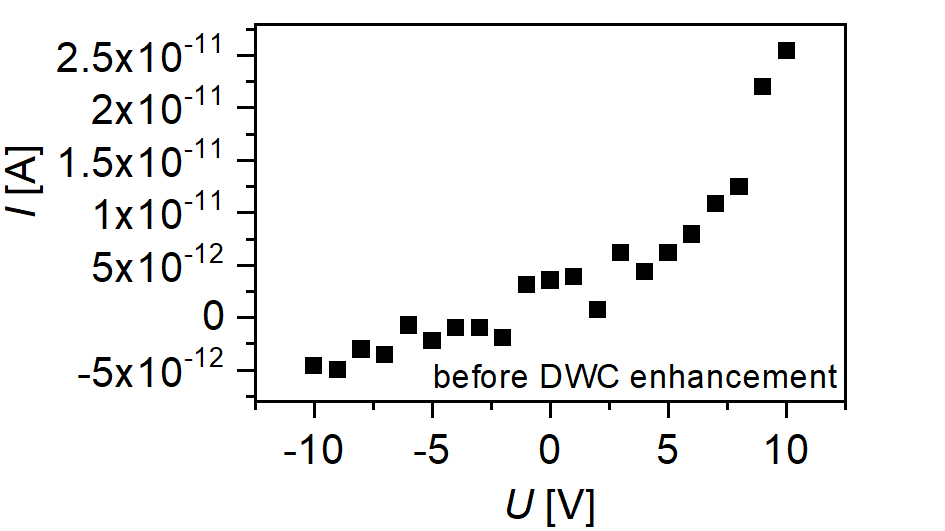}
		\caption{}
		\label{fig:subfig_b}
	\end{subfigure}
	\hfill
	\begin{subfigure}{0.37\textwidth}
		\centering
		\includegraphics{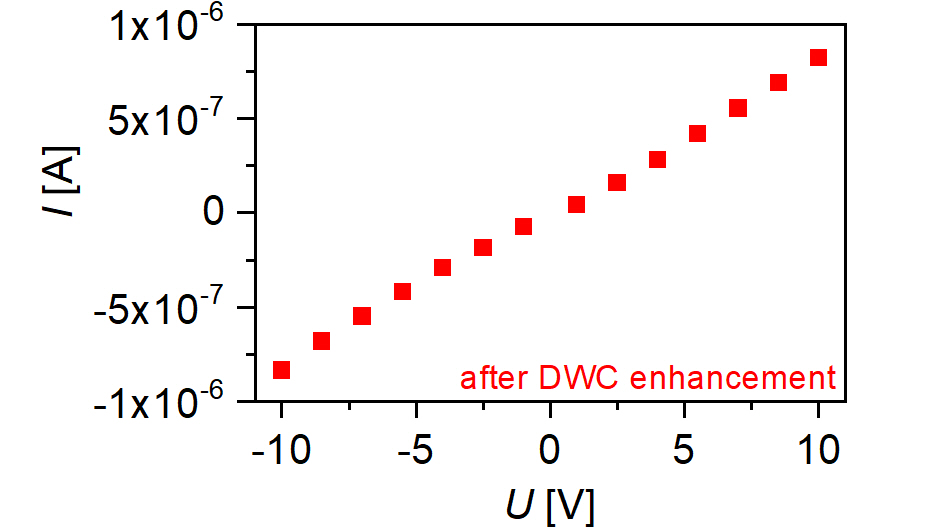}
		\caption{}
		\label{fig:subfig_c}
	\end{subfigure}
	\caption{\textbf{LN3-a:} (a) Domain image; (b) current-voltage characteristics after poling,
		and (c) after DWC enhancement.} % Overall caption for both subfigures
	\label{fig:4-1} % Label for the entire figure
\end{figure}

Following the UV-assisted poling process, images of the initial hexagonal domains are captured using a polarizing microscope. The domain images for LN1-a, LN2-a, LN2-b and LN3-a taken directly after
poling, are shown in Figs. \ref{fig:subfig_a1}, \ref{fig:subfig_a2},  \ref{fig:subfig_a3}, and \ref{fig:subfig_a}. The current-voltage (\IV) characteristics of these domain walls (DWs) are measured using an electrometer, and presented in Figs. \ref{fig:subfig_b1}, \ref{fig:subfig_b2}, \ref{fig:subfig_b3}, and \ref{fig:subfig_b}. After domain wall conductivity (DWC) enhancement, the \IV-characteristics shown in Figs. \ref{fig:subfig_c1}, \ref{fig:subfig_c2}, \ref{fig:subfig_c3}, and \ref{fig:subfig_c} result.

After poling, \IV-characteristics of domain LN1-a (Fig. \ref{fig:subfig_b1}) and LN2-b (Fig. \ref{fig:subfig_b3}) exhibit a linear relationship, thereby indicating ohmic behavior. However, after poling, the \IV-curve obtained for domains LN2-a (Fig. \ref{fig:subfig_b2}) and LN3-a  (Fig. \ref{fig:subfig_b}) exhibits a nonlinear behavior pointing to diode-like contribution. All experiments presented in section S1 are conducted at room temperature.

\newpage

\textbf{Supplement S2: Micro-impedance setup.}

\begin{figure}[H]%[htbp]
	\centering
	\includegraphics[scale=0.6]{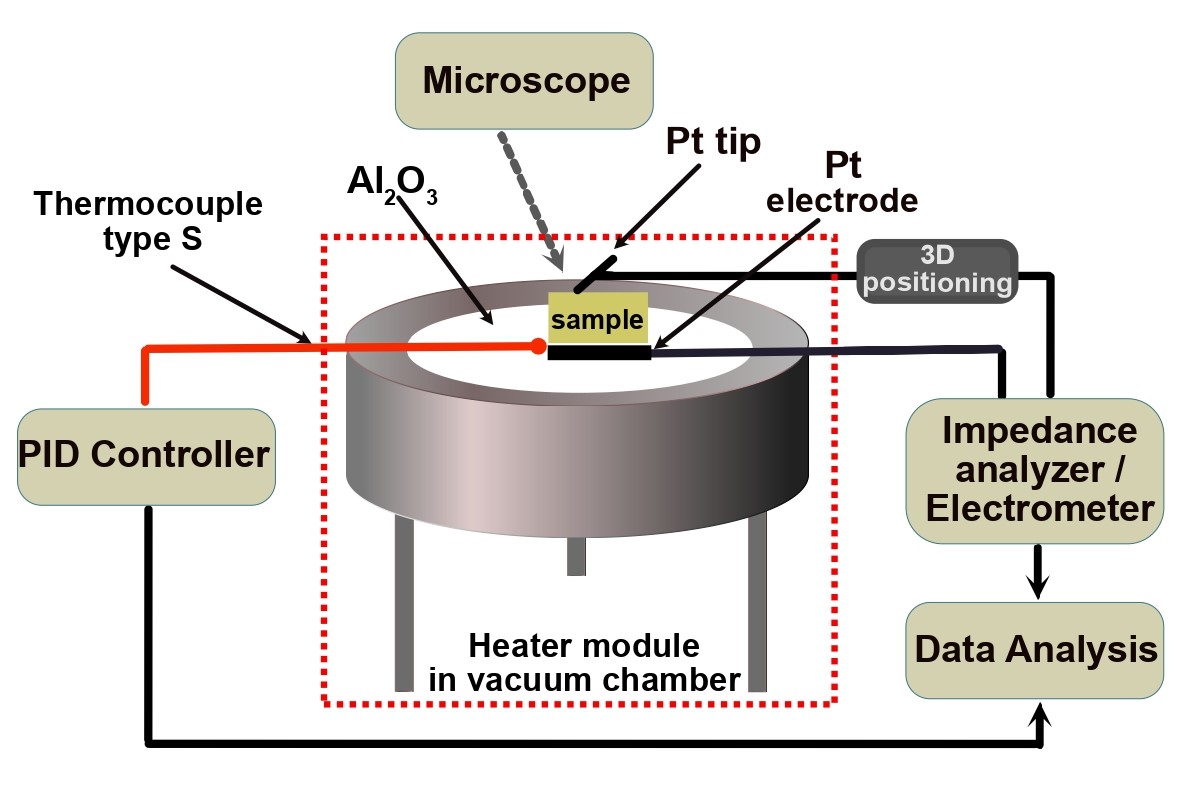}% Adjust scale if needed
	\caption{Micro-impedance setup  for high temperature measurement.}
	\label{fig:5}
\end{figure}

To measure domain wall currents (DWCs at high temperatures, a micro-impedance setup is employed. This setup includes a heater module (UHV Design, UK) and precise positioning screws for the top-contact platinum tip, as depicted in Fig. \ref{fig:5}. In this setup, the sample lays on a Pt electrode, both of which are placed on a ceramic plate atop the heater module. The heating element of the module is made of high-density graphite coated with pyrolytic graphite. The entire setup, including the sample, electrodes, and heater module, is enclosed in a vacuum chamber maintained at a total pressure of approximately 2\,Pa. Inside the chamber, a type S thermocouple is positioned within the ceramic plate on top of the heater module, situated about 1-2\,mm away from the sample.

This chamber is connected to either an impedance analyzer or, in this specific case here, to an highly sensitive amperometer (Keithley 6517B electrometer). Current-voltage sweeps ranging from -10 to +10 V are applied, with a step size of 1.5\,V. Each sweep takes around 14\,s to complete. This method enables data collection at different temperatures, with a heating rate of 1\,K/min. The setup reaches a maximum temperature of 400$^\circ$C.

\newpage
\textbf{Supplement S3: \IV-curve across various temperatures, up to 375$^\circ$C post-enhancement.}

\begin{figure}[H]%[htbp]
	\centering
	\begin{subfigure}{0.32\textwidth}
		\centering
		\includegraphics[scale=1.25]{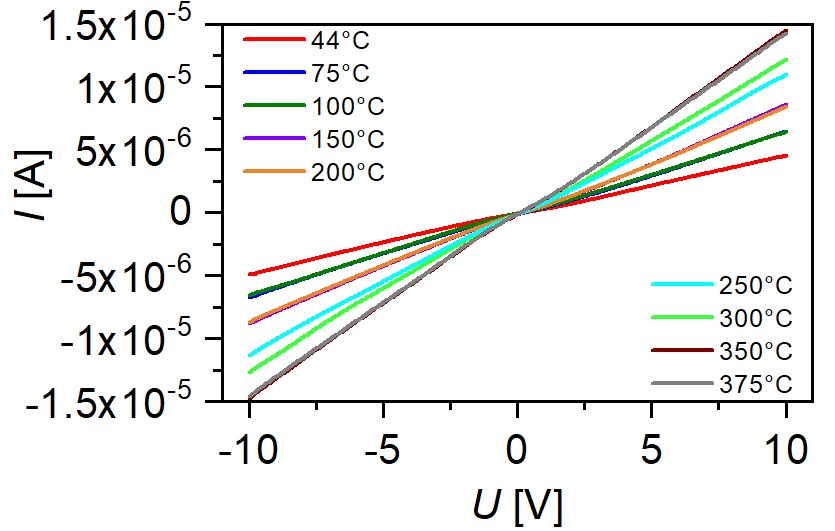}
		\caption{\textbf{LN1-a}}
		\label{fig:subfig_a4}
	\end{subfigure}
	\hfill
	\begin{subfigure}{0.5\textwidth}
		\centering
		\includegraphics[scale=1.25]{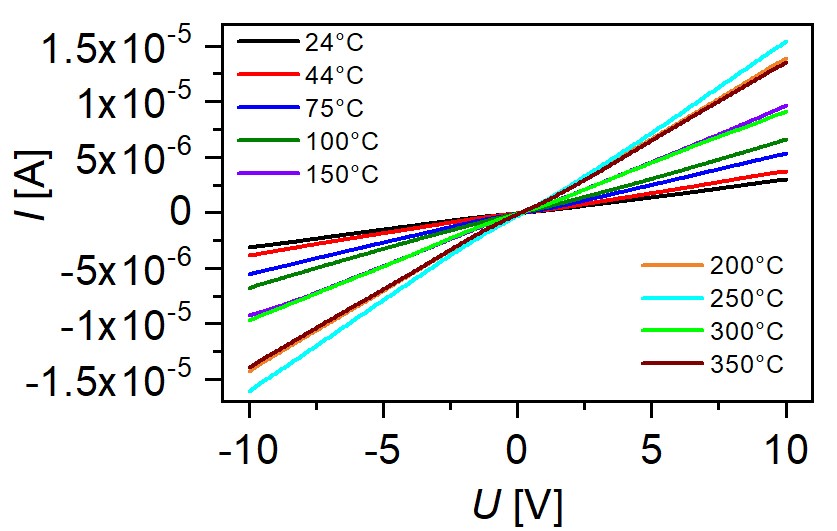}
		\caption{\textbf{LN2-a}}
		\label{fig:subfig_b4}
	\end{subfigure}
	%\hfill
	\\
	\begin{subfigure}{0.5\textwidth}
		\centering
		\includegraphics[scale=1.25]{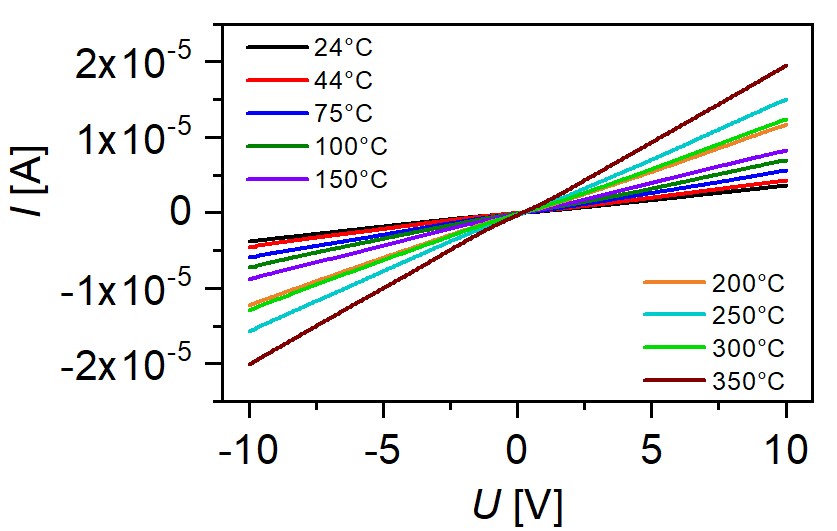}
		\caption{\textbf{LN2-b}} % Caption for the first subfigure (you can add text here)
		\label{fig:subfig_c4}
	\end{subfigure}
	\caption{\IV-curve across various temperatures, up to 375$^\circ$C. Note the ohmic behavior
		of all \IV-curves, being independent on temperature or domain size.} % Overall caption for both subfigures
	\label{fig:6} % Label for the entire figure
\end{figure}

Following the enhancement of DWC, the \IV-curves of domains LN1-a, LN2-a, and LN2-b exhibit ohmic behavior. Figs. \ref{fig:subfig_a4}, \ref{fig:subfig_b4}, and \ref{fig:subfig_c4} illustrate the \IV-curves at different temperatures, reaching up to 375$^\circ$C. 

\newpage
\textbf{Supplement S4: Arrhenius-type plot of the current of LN2-a and LN2-b in a temperature range from room temperature up to 400$^\circ$C. }

\begin{figure}[htbp]
	\centering
	\includegraphics[scale=1.2]{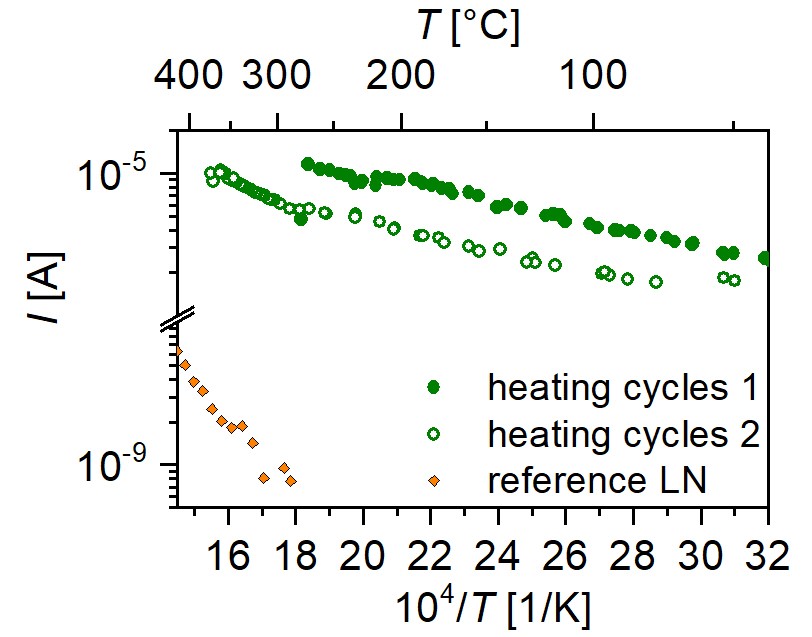}% Adjust scale if needed
	\caption{ Temperature-dependent current of LN2-a at +7~V for both the initial and subsequent heating cycle.   }
	\label{fig:7}
\end{figure}

\begin{figure}[htbp]
	\centering
	\includegraphics[scale=1.2]{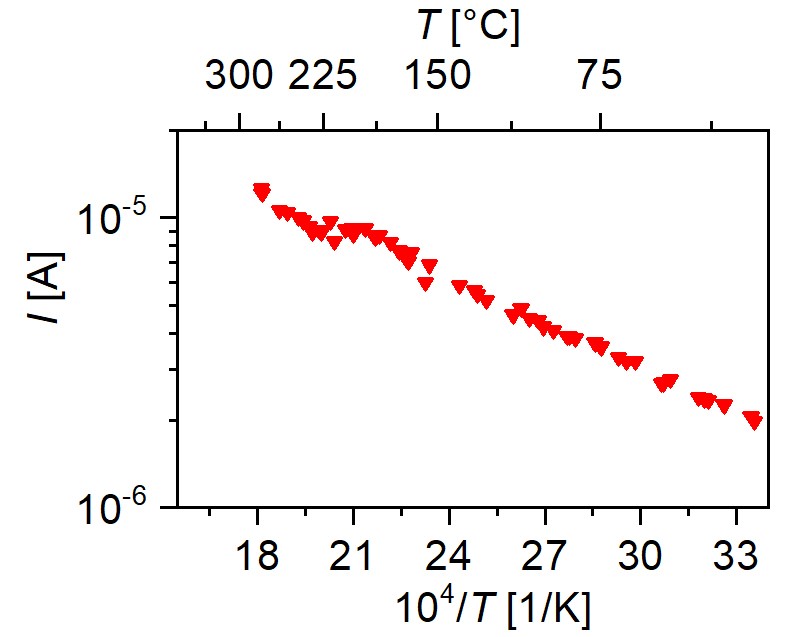}% Adjust scale if needed
	\caption{ Temperature-dependent current of LN2-b at +7~V.}
	\label{fig:8}
\end{figure}

Figures \ref{fig:7} and \ref{fig:8} depict the DWC measured at a bias voltage of +7~V for LN2-a and LN2-b, respectively, presented in Arrhenius-type plots. Notably, both figures illustrate a nearly linear increase in DWC, although with varying slopes. This observation also suggests the presence of thermally-activated transport processes characterized by different activation energies. 

\newpage
\textbf{Supplement S5: Atomistic calculations.}

Atomistic calculations as described in the main article were performed on supercells of different symmetries. X and Y DWs are parallel to the crystal X and Y axis, and thus parallel to the (10$\overline{1}$0) and (11$\overline{2}$0) LN lattice planes, respectively. Both types of DWs can be modelled by means of orthorhombic supercells. A 1$\times$6$\times$1 and a 6$\times$1$\times$1 repetition of the orthogonal unit cell are employed to model X and Y DWs, respectively. Within these supercells, two X and Y DWs are separated by about 25\,{\AA} and by about 15\,{\AA}.  6$\times$3$\times$2 and 4$\times$3$\times$2 Monkhorst-Pack k-point meshes are used for the energy integration of the supercells modelling X and Y DWs, respectively. 
H2H and T2T DWs are modelled by means of a 1$\times$1$\times$12 repetition of the hexagonal unit cell and a 4$\times$4$\times$1  Monkhorst-Pack k-point mesh. Tests with larger and smaller supercells have ben performed in all cases to ensure numerically converged results with respect to the DW separation.
Atomic structures are relaxed untill the Helman-Feynman forces acting on the single atoms are lower than a threshold of 10$^{-3}$\,eV/{\AA}.

\begin{figure}[htbp]
	\centering
	\includegraphics[scale=1.0]{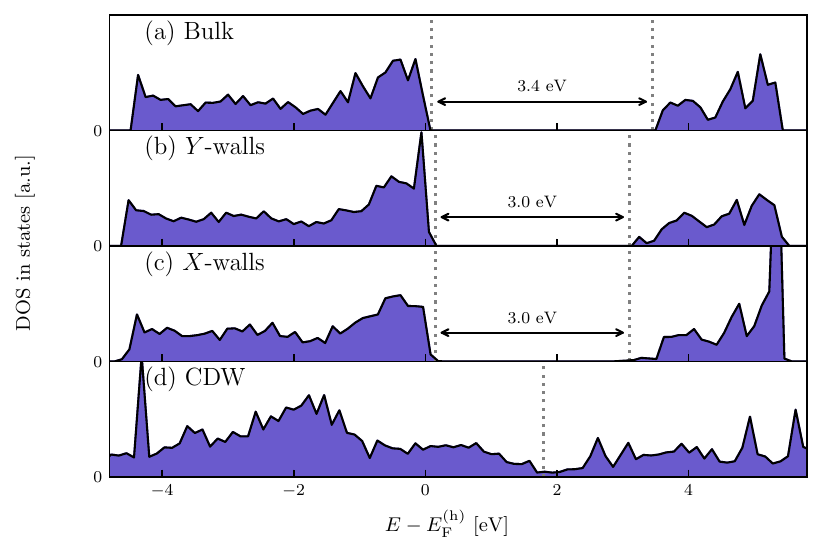}% Adjust scale if needed
	\caption{ DFT-calculated total DOS of LN for the crystal bulk (a) and different wall types: (b) Y walls, (c) X walls, (d) H2H and T2T walls, labeled as charged domain walls (CDW).}
	\label{fig:9}
\end{figure}

Figure \ref{fig:9} shows the calculated density of states (DOS) of different structures. Panel (a) represents the DOS of LN buk. The fundamental gap of about 3.4\,eV is clearly visible. Panels (b) and (c) show the DOS of the supercells modelling Y and X DWs, respectively. In both cases the electronic gap shrinks to about 3.0\,eV. Panel (d) illustrates the DOS calculated for the supercell modelling the fully charged domain walls (CDW), for which valence and conduction bands are merged.

\end{document}